PLANETARY FORMATION SCENARIOS REVISTIED:
CORE-ACCRETION VERSUS DISK INSTABILITY

Matsuo, T., Shibai, H., AND Ootsubo, T.,
Nagoya University, Furo-cho, Chikusa-ku, Nagoya, 464-8602, Japan;
matsuo@u.phys.nagoya-u.ac.jp
AND
Tamura, M.
National Astronomical Observatory, Osawa, Mitaka, Tokyo, 181-8588, Japan;

## ABSTRACT

The core-accretion and disk instability models have so far been used to explain planetary formation. These models have different conditions, such as planet mass, disk mass, and metallicity for formation of gas giants. The core-accretion model has a metallicity condition ([Fe/H] > −1.17 in the case of G-type stars), and the mass of planets formed is less than 6 times that of the Jupiter mass $M_J$. On the other hand, the disk instability model does not have the metallicity condition, but requires the disk to be 15 times more massive compared to the minimum mass solar nebulae model. The mass of planets formed is more than $2M_J$. These results are compared to the 161 detected planets for each spectral type of the central stars. The results show that 90% of the detected planets are consistent with the core-accretion model regardless of the spectral type. The remaining 10% are not in the region explained by the core-accretion model, but are explained by the disk instability model. We derived the metallicity dependence of the formation probability of gas giants for the core-accretion model. Comparing the result with the observed fraction having gas giants, they are found to be consistent. On the other hand, the observation cannot be explained by the disk instability model, because the condition for gas giant formation is independent of the metallicity. Consequently, most of planets detected so far are thought to have been formed by the core-accretion process, and the rest by the disk instability process.

# 1. INTRODUCTION

The planetary formation theory has been developed mainly for the solar system. There are two representative models, the core-accretion scenario (e.g., Safronov 1969; Goldreich & Ward 1973; Hayashi, Nakazawa, & Nakagawa 1985; Pollack et al. 1996) and the disk instability scenario (e.g., Kuiper 1951; Cameron 1978). According to the core-accretion scenario, a heavy element core is built by the accretion of planetesimals. As the core grows, its ability to accrete gas from the surrounding disk increases. When the core is sufficiently massive, rapid gas accretion occurs onto the core and a gas giant is formed. The same basic mechanism may also form the terrestrial planets. However, the problem with this scenario is that the time taken for gas giant formation is close to the upper limit estimated for the gas depletion timescale of the disks (Haisch, Lada, & Lada 2001). On the other hand, according to the disk instability scenario, if a disk is sufficiently massive, the disk fragments into a dense core. Such clumps can contract to form giant gaseous protoplanets in several hundred years. Gas giants are quickly formed before the gas in the disk depletes. However, gas giants formed due to disk instability are assumed to have a solar abundance ratio, which is inconsistent with the indication that the envelopes of Jupiter and Saturn are enriched with heavy elements (Saumon et al. 1995; Young 2003). On the other hand, gas giants formed through core accretion are enriched due to accretion of planetesimals. Therefore, the core-accretion scenario has been widely accepted as the standard model for the formation of the solar system.

Since the first discovery of an extrasolar planet by Mayor & Queloz (1995), 190 extrasolar planets have so far been detected. Fischer & Valenti (2005) and Santos et al. (2004) found that the fraction of gas giant detection increases as the metallicity of the central star increases. This fact is explained by the primordial origin of metallicity (Gonzalez 1997). The amount of core material in a circumstellar disk is proportional to the metallicity of the central star, because a central star and its circumstellar disk are born from a mother cloud core. Ida & Lin (2004b) and Kornet et al. (2005) theoretically derived the probability of gas giant formation based on the primordial origin hypothesis of the metallicity. Further, Robinson et al. (2006) found that silicon and nickel are significantly enriched in planet host stars. This observational fact supports the primordial origin hypothesis.

Ida & Lin (2004a) and Alibert et al. (2005) theoretically predicted the mass and the semi-major axis of planets formed through the core-accretion process, including migration, disc evolution, and gap formation. They compared the final masses and semi-major axes distribution generated in their simulations with that of planets discovered so far. Ida & Lin (2004a) concluded that the results of the simulation are consistent with the observational facts. In addition, Alibert et al. (2005) theoretically found that the migration significantly reduces the timescale of gas giants formation, because the migration effectively prevents the gas depletion of the feeding zone. They showed that the gas giants can be formed within the disk lifetime. Therefore, the core-accretion scenario has broadly been accepted not only for the planet formation process in the solar system but also in extrasolar planetary systems.

On the other hand, Fukagawa et al. (2004) found a spiral structure in the circumstellar disk of AB Aurigae by high resolution imaging with the stellar coronagraph of the Subaru telescope (CIAO/Subaru), and concluded that this structure was formed due to weak gravitational instability. In addition to this observational evidence, Boss (1997) indicated that the timescale for gas giant formation in the disk instability process (several hundred years) is much shorter than that in the core-accretion process (several million years) (e.g., Pollack et al. 1996). Therefore, once the condition for disk instability is satisfied, gas giants will be formed by the disk instability process. However, observational evidence on gas giant formation by the disk instability process has not been presented.

On the other hand, Laughlin & Bodenheimer (1994), Boss (1997, 2002), Mayer et al. (2002, 2004), Johnson & Gammie (2003), Pickett et al. (2003), and Majia et al. (2005) describe the evolution of gravitationally unstable gaseous protoplanetary disks using two- and three-dimensional hydrodynamic simulations. Gammie (2001) and Rice et al. (2003) found that the disks fragment into dense protoplanetary clumps if the cooling time is shorter than a few orbital periods. Boss (2002) carried out three-dimensional hydrodynamic disk simulations with radiative cooling and found that the disks fragment into dense clumps for all metallicities in the range $0.1 < f_d / f_g < 10$. On the other hand, Cai et al. (2006) showed that, since the cooling time of the disks is longer than the timescale of the disk fragment, disk fragmentation into a dense core cannot occur for all metallicities in the range $\frac{1}{4} < \frac{f_d}{f_g} < 2.0$. As mentioned by Cai et al. (2006), the apparent disagreement between these results could come from the differences in cooling timescales due to the differences in techniques and assumptions, such as artificial viscosity, opacities, equations of state, initial disk models, and perturbations. This paper assumes that disk fragmentation occurs.

In this paper, we examine the conditions for gas giant formation in two models with respect to metallicity dependence: The lower limits for gas giants formed by the two processes are derived. We also derive the mass of the planets that are thought to be formed through two processes: The maximum mass for gas giants formed by the core-accretion process and the minimum mass for gas giants formed by disk instability are derived. We compare these results with the observational results in order to examine whether or not all planets are thought to be formed by the core-accretion scenario, and whether gas giant formation by disk instability actually took place.

In section 2, we derive the conditions for stellar metallicity and planet mass in the core-accretion model. In section 3, we derive the conditions for the disk instability model. In section 4, we compare the planets detected so far with the derived conditions. In section 5, we discuss the possibility of planetary formation by the core-accretion and disk instability scenario.

## 2. CORE-ACCRETION MODEL

In this section, we derive the conditions for planet mass and metallicity of the protoplanetary disks for gas giant formation in the core-accretion model. We assume that the host stellar mass is 1 solar mass, $M_* = 1M_{sun}$, and the temperature distribution of the disk is given by the Hayashi model (Hayashi 1981) unless otherwise denoted. The gas surface density is 0.01 to 10 times the minimum mass solar nebula model (hereafter, the MMSN model).

### 2.1. *Condition For Metallicity*

We derive the range of disk metallicity for gas giant formation by the core-accretion process. According to Hayashi et al. (1985), the conditions for gas giant formation are given as follows:

1. The core isolation mass, $M_{c,iso}$, is larger than the critical core mass, $M_{c,crit}$.

$$M_{c,iso} \geq M_{c,crit}. \tag{1}$$

The core isolation mass is defined as the total mass of the planetesimals existing in the feeding zone of the core, and the critical core mass is defined such that above that value the ambient gas cannot be supported by the hydrostatic pressure and begins to accrete rapidly onto the core.

2. The core mass increases over the critical core mass before the disk gas dissipates,

$$\tau_{c,acc} \leq \tau_{disk}, \tag{2}$$

where $\tau_{c,acc}$ is defined as the time to acquire the critical core mass and $\tau_{disk}$ is the gas dissipation timescale of the disk.

According to Kokubo & Ida (2002), the core isolation mass is given by

$$M_{c,iso} \approx 0.16 \eta_{ice}^{\frac{3}{2}} f_d^{\frac{3}{2}} \left(\frac{a}{1AU}\right)^3 \left(\frac{M_*}{M_{sun}}\right)^{-\frac{1}{2}} M_\oplus, \tag{3}$$

where $f_d$ is the scaling factor for the dust surface density of the MMSN model, and $M_\oplus$ is the Earth mass. $\eta_{ice}$ is the scaling factor for the dust surface density in icy conditions.

$$\eta_{ice} = 1 \quad (a \leq a_{ice}) \tag{4}$$

$$\eta_{ice} = 4.2 \quad (a \geq a_{ice}), \tag{5}$$

where $a_{ice}$ is the semi-major axis of the snow line.

Mizuno (1980) and Ikoma, Nakazawa, & Emori (2000) proposed that the critical core mass is 4 to 30 times that of the Earth's mass. We adopt the smallest value (4 times) as the critical core mass in order to obtain the weakest condition. Then, the first condition (1) is

$$a \geq a_{tg} \approx 4.1 \times f_d^{-2} \left(\frac{\eta_{ice}}{4.2}\right)^{-2} \left(\frac{M_{c,crit}}{4M_\oplus}\right)^{\frac{4}{3}} \left(\frac{M}{M_{sun}}\right)^{\frac{2}{3}} AU, \tag{6}$$

where $a_{tg}$ is the minimum semi-major axis in which gas giants can be formed.

Ida & Lin (2004a) derived the core accretion time as

$$\tau_{c,acc} = \frac{M_c}{\dot{M}_c} \approx 1.2\times10^5 \eta_{ice}^{-1} f_d^{-1} f_g^{-\frac{2}{5}} \left(\frac{a}{1AU}\right)^{\frac{27}{10}} \left(\frac{M_c}{M_\oplus}\right)^{\frac{1}{3}} \left(\frac{M_*}{M_{sun}}\right)^{-\frac{1}{6}} yrs. \qquad (7)$$

Assuming that the gas depletion timescale of the disks is given as one million years based on the observation results of Wyatt et al. (2003), the second condition is replaced with

$$a \le a_{gi} \approx 7.4\times f_g^{\frac{4}{27}} f_d^{\frac{10}{27}} \left(\frac{M_c}{4M_\oplus}\right)^{-\frac{10}{81}} \left(\frac{\eta_{ice}}{4.2}\right)^{\frac{10}{27}} \left(\frac{\tau_{disk}}{10^7 yrs}\right)^{\frac{10}{27}} \left(\frac{M}{M_{sun}}\right)^{\frac{5}{81}} AU, \qquad (8)$$

where $a_{gi}$ is the largest semi-major axis in which gas giants can be formed, and $f_g$ is the scaling factor for the gas surface density of the MMSN model.

Here, according to the primordial origin hypothesis, we assume that the metallicity of a star, $\left[Fe/H\right]_*$, is equal to that of its disk, $Z_{disk}$, that is, $\left[Fe/H\right]_* \approx Z_{disk} = \log_{10}\left(\frac{f_d}{f_g}\right)$. Therefore, equation (6) is written as

$$a \ge a_{tg} \approx 4.1\times f_g^{-2}\cdot 10^{-2Z_{disk}} \left(\frac{\eta_{ice}}{4.2}\right)^{-2} \left(\frac{M_{c,crit}}{4M_\oplus}\right)^{\frac{4}{3}} \left(\frac{M_*}{M_{sun}}\right)^{\frac{2}{3}} AU,. \qquad (9)$$

and equation (8) can be written as

$$a \le a_{gi} \approx 7.4\times f_g^{\frac{14}{27}} 10^{\frac{10}{27}Z_{disk}} \left(\frac{M_c}{4M_\oplus}\right)^{-\frac{10}{81}} \left(\frac{\eta_{ice}}{4.2}\right)^{\frac{10}{27}} \left(\frac{\tau_{disk}}{10^7 yrs}\right)^{\frac{10}{27}} \left(\frac{M_*}{M_{sun}}\right)^{\frac{5}{81}} AU. \qquad (10)$$

Therefore, the condition of disk metallicity for gas giant formation is

$$\left[Fe/H\right] \ge \left[Fe/H\right]_{crit} \approx -0.11 + \log\left\{\left(\frac{f_g}{1}\right)^{-\frac{17}{16}} \left(\frac{\eta_{ice}}{4.2}\right)^{-1} \left(\frac{\tau_{disk}}{10^7 yrs}\right)^{-\frac{5}{32}} \left(\frac{M_{c,crit}}{4M_\oplus}\right)^{\frac{59}{96}} \left(\frac{M_*}{M_{sun}}\right)^{\frac{49}{192}}\right\}$$

(11)

In the case of $f_g$ = 5 and 10, the lower limits for metallicity $\left[{}^{Fe}\!/_{H}\right]_{crit}$ are −0.85 and −1.17, respectively. These results are based on the assumption of the static power-law disk. Figure 1 represents the condition of disk metallicity for gas giant formation. The lower limits for the metallicity value are indicated by the solid lines in Figure 1.

### 2.2. *Upper Limit For Planet Mass*

The maximum mass for a gas giant formed by core accretion is derived. Hayashi et al. (1985) showed that, in an inviscid disk the mass of a gas giant is considered to be approximately given by the total mass of the gas in its feeding zone. On the other hand, Lin & Papaloizou (1993) indicated that in a viscously evolving disk, gas can continuously diffuse into the feeding zone due to the pressure gradient and viscous diffusion, and gas accretion onto the core is terminated when gravitational scattering is equal to these effects. Cabot et al. (1987) showed theoretically that turbulent efficiency ranges from 0.01 to 0.0001. Based on these studies, Ida & Lin (2004a) stated that, because gas accretion onto the core is not controlled by viscous diffusion, but mainly by the pressure gradient except for some very large $\alpha$ ($\alpha$< 0.001), gas inflow onto the core would actually be stopped when gas pressure and gravitational scattering are comparable and the radius of the gap formed is about scale height *h*. The final mass of the gas giant ($M_{g,th}$) is given as follows:

$$M_{g,th} \approx 0.38 \left( \frac{a}{1AU} \right)^{\frac{3}{4}} \left( \frac{M_*}{M_{sun}} \right) M_J , \tag{12}$$

where $h^H$ is the scale height when the temperature distribution is obtained from the Hayashi model (Hayashi 1981). On the other hand, in the case of very large $\alpha$ (0.01 > $\alpha$ > 0.001), gas accretion onto the core is controlled by viscous diffusion. Ida & Lin (2004) derived an equation for the final mass of the gas giant ($M_{g,vis}$), in which viscous diffusion is associated with gravitational scattering as follows:

$$M_{g,vis} \approx 0.94\left(\frac{\alpha}{0.01}\right)\left(\frac{a}{1AU}\right)^{\frac{1}{2}}\left(\frac{M_*}{M_{sun}}\right)M_J . \quad (13)$$

Equations (12) and (13) depend on semi-major axis ($a$). When $a$ is a maximum value, $a \approx a_{gi}$, $M_{g,th}$ and $M_{g,vis}$ take the maximum value,

$$M_{p,\max 1} = M_{g,th}\left(a = a_{gi}\right) \approx 1.70\times 10^{\frac{5}{18}Z_{disk}}\left(\frac{f_g}{1}\right)^{\frac{7}{18}}\left(\frac{h}{h^H}\right)^{3}\left(\frac{\tau_{disk}}{10^7 yrs}\right)^{\frac{5}{18}}\left(\frac{M_*}{M_{sun}}\right)^{\frac{113}{108}}M_J , \quad (14)$$

and

$$M_{p,\max 2} = M_{g,vis}\left(a = a_{gi}\right) \approx 2.56\times 10^{\frac{5}{27}Z_{disk}}\left(\frac{f_g}{1}\right)^{\frac{7}{27}}\left(\frac{\alpha}{0.01}\right)\left(\frac{\tau_{disk}}{10^7 yrs}\right)^{\frac{5}{27}}\left(\frac{M_*}{M_{sun}}\right)^{\frac{167}{162}}M_J , \quad (15)$$

respectively. Equations (14) and (15) are shown as solid and dashed lines in Figure 2, respectively. From Figure 2 it can be concluded that the final planet masses given by equations (12) and (13) are approximately comparable.

The above assumption is very sensitive to scale height, $h$. $h$ is maximum in the disk model whose limit is optically thin (Hayashi model), and the final mass becomes the maximum value. In contrast, in the case of an optically thick disk (Kusaka, Nakamoto & Hayashi 1970), the disk temperature is one-third that of a thin disk. Therefore, the massive planets above the solid and dash lines in Figure 2 cannot be formed by the core-accretion process, right after the gap is formed.

After the gap has formed, gas around the gap may accrete onto the protoplanet. D'Anglelo, Henning, & Kley (2002), and D'Anglelo, Kley, & Henning (2003) investigated the evolution of protoplanets with different masses embedded in an accretion disk, using two- and three-dimensional hydrodynamic simulations. As a result, they derived an equation for the gas accretion onto the protoplanets as

$$\log\left(\frac{\dot{M}_p}{M_{\oplus} yr^{-1}}\right) \approx -18.47 - 9.25\log q - 1.266\left(\log q\right)^2 , \quad (16)$$

where $q$ is the planet-to-star mass ratio. Haisch, Lada, & Lada (2001) and Wyatt et al. (2003) revealed that the gas depletion timescale of the disks ranges from 1 Myr to 10 Myrs based on the young stellar object (YSO) of submillimeter observation. Based on these results, even when the disk surface density is 10 times the MMSN model ($f_g = 10$), which is more for a passive disk, the planet mass $4.5\,M_J$ in which the gap is formed increases to a maximum of $6\,M_J$ after 10 million years. This is the maximum mass of the planet formed through core accretion. Figure 2 shows the maximum planet mass. In case that the type-Ⅰ and type-Ⅱ migration is not considered, the range in which a gas giant is thought to be formed through core accretion is indicated by the inside of the boundary line in Figure 3.

## 3. DISK INSTABILITY MODEL

According to the previous work on the disk instability scenario (e.g., Kuiper 1951; Laghlin & Bodenheimer 1994; Boss 1997; Gemmie 2001; Mayer et al. 2002, 2004; Johnson & Gammie 2003, Pickett et al. 2003, Rice et al. 2003), if a disk is gravitationally unstable and the cooling time is shorter than a few orbital periods, the disk fragments into a dense protoplanetary clump. Direct gas giant formation due to disk instability may occur within a timescale of several hundred years.

In the early phase of disk evolution, Bachiller (1996) mentioned that the disks are very massive and optically thick. Therefore, we adopt the disk temperature distribution as in the optically thick Kusaka model (Kusaka et al. 1970). The disk surface density is 10 to 30 times that of the MMSN model ($10 < f_g < 30$).

### 3.1. *Lower Limit For Gas Giant Formed By Disk Instability*

We simply derive the minimum planet mass that can be formed by the disk instability process, $M_{G,\min}$, assuming that disk instability is the result of local fragmentation via ring formation (m=0) in a WKB disk. When the disk instability occurs, the entire disk is thought to be global instable. In the disk, a perturbation of a spatial scale shorter than the critical wavelength cannot grow up. On the other hand, the perturbation of a spatial scale larger than the critical wavelength can form a local fragment. Therefore, the mass of the fragment whose size equals to the critical wavelength is thought to be the minimum planet mass through the disk instability. On the other hand, the assumption that each fragment forms single planet is still questionable. Thus, the minimum planet mass derived here is uncertain in this mean.

The condition for gravitational instability of the disks is that the Toomre-Q value should be less than or equal to 1, that is,

$$Q = \frac{\Sigma_{crit}}{\Sigma_g} \approx \frac{c_s \Omega_k}{\pi G \Sigma_g} < 1, \tag{17}$$

where $\Sigma_{crit}$ is the critical surface density, $\Sigma_g$ is the gas surface density of the disks, $c_s$ is the isothermal sonic velocity, and $\Omega_k$ is the keplerian angular velocity. In this case, the minimum mass of the planets is given by

$$M_G \sim \pi \Sigma_g \left( \frac{2\pi}{|k_{crit}|} \right)^2, \tag{18}$$

where $k_{crit}$ represents the critical wavelength number, that is,

$$k_{crit} \sim \frac{\pi G \Sigma_{crit}}{c_s^2} \sim \frac{\Omega_K}{c_s}. \tag{19}$$

The planet mass and the condition for disk instability are determined by the surface density and temperature of the disks. According to Kusaka et al. (1970), the temperature distribution of an optically-thick disk is described as

$$T^4 = \left( \frac{L_* R_*}{6\pi^2 \sigma_{SB}} \right) a^{-3} + \left( \frac{L_*}{28\pi \sigma_{SB}} \right)^{\frac{8}{7}} \left( \frac{2k_B}{\mu m_H G M_*} \right)^{\frac{4}{7}} a^{-\frac{12}{7}}, \tag{20}$$

where $L_*$ is the stellar luminosity, $R_*$ is the stellar radius, and $\mu$ is the mean molecular weight. The mean molecular weight is assumed to be 2.4, because the disk gas is mainly composed of molecular hydrogen (80%) and helium (20%).

Figure 4 shows the dependence of the minimum planet mass formed by disk instability on the semi-major axis. Less massive planets cannot be formed in the region with larger semi-major axes. The reason is that the range where gas can contract due to self-gravitation, $a_{grav}$, is inverse of the critical wavelength number, that is, $a_{grav} \sim \frac{1}{k_{crit}} \propto a^{\frac{3}{2}} \cdot T^{\frac{1}{2}} \cdot M_*^{-\frac{1}{2}}$. In addition, the range in which gravitational instability of the disk takes place is determined by the disk temperature and the gas surface density, as expected from the Toomre-Q value, $Q \propto f_g^{-1} \cdot T^{\frac{1}{2}} \cdot M_*^{\frac{1}{2}}$. As shown in Figure 4, the region where gravitational instability of the disk occurs, widens inwardly as the gas surface density of the disks increases. The reason is that the nearer the host star, the higher the gas pressure, and the stronger the gravitation of the host star. Therefore, the planets of minimum mass are formed in the place where the Toomre-Q value equals 1. For example, if the gas surface density is 20 times that of the MMSN ($f_g = 15$), the semi-major axis where the Toomre-Q value is equal to 1 is 7.5 AU, and the minimum planet mass is $1.76 M_J$. In addition, when the gas surface density is 20 and 30 times that of the MMSN ($f_g = 20,30$), the region where gravitational instability occurs is outside of 2.1 and 0.5 AU, and the minimum mass would be $0.21 M_J$ and $0.61 M_J$, respectively. On the other hand, it is difficult to theoretically estimate the maximum mass of the planets formed by disk instability.

Next, we discuss the metallicity dependence of gravitational instability of disks. Boss (2002) carried out three-dimensional hydrodynamic disk simulations with radiative cooling and found that the disks fragment into dense clumps for all metallicities over the range $0.1 < f_d / f_g < 10$. On the other hand, Cai et al. (2006) showed that the cooling time of the disks is too long for disk fragments to occur for all metallicities over the range $1/4 < f_d / f_g < 2.0$. As mentioned by Cai et al. (2006), the apparent disagreement between these results could come from the difference in cooling timescales due to differences in techniques and assumptions, such as artificial viscosity, opacities, equations of state, initial disk models, and perturbations.

Figure 5 represents the upper limits for the mass of gas giants formed by disk instability in the case of G-type stars. The region where gas giants are thought to be formed by disk instability is the upper side of the boundary line.

# 4. SAMPLES

## 4.1. *Samples*

The samples in this study are the 161 planets detected so far by the radial velocity survey. The major data sources for metallicity of the host stars are Santos et al. (2004; 2005), Laws et al. (2003), and Bonfils et al. (2005) for M-type stars. The other sources are those derived from spectroscopic data measured during radial velocity observation. We use the weighted mean of all available data for sources that have two or more independent data for metallicity. The observational data and its references are compiled in Table 1-1. In addition, we refer to the data in "Extrasolar Planets Encyclopedia" for the planet mass values.

## 4.2. *Correlation Between Metallicity Of The Host Stars And The Planet Mass*

We compare the samples with the boundaries of the two models derived in the previous sections for stars of each spectral type, because the boundaries are sensitive to central stellar mass. In addition, the spectral types of subgiants or giants are different from those in the main sequence phase. Therefore, it is necessary to remove subgiants or giants from all samples. Figure 6a, b, and c represent the correlation between metallicity of the host stars and the planet mass for all samples, G-type stars, and G-type dwarfs in the main sequence phase, respectively. As shown in Figure 6a, the distribution of samples in each case is almost the same. Therefore, we use the samples refined to those of only dwarfs, since the discussion is independent of each sample.

For the comparison of the conditions for gas giant formation in each model with samples (105), we know that these planets were formed by either model. Figure 7a, b, c, and d show the comparison of samples in the region where gas giants can be formed by two planetary formation mechanisms for F-type stars, G-type stars, K-type stars, and M-type stars, respectively. For the derivation of theoretical boundaries, the stellar masses of F-type, G-type, K-type, and M-type stars are $1.3\,M_{sun}$, $1.0\,M_{sun}$, $0.7M_{sun}$, and $0.4\,M_{sun}$, respectively. In addition, the relationship between stellar mass and luminosity or radius is given by $L_* \propto M_*^{3.5}$, $R_* \propto M_*$, respectively.

## 5. DISCUSSION

### 5.1. *Observational Facts That Back Up The Core-accretion Scenario*

In the case of G-type dwarfs, as shown in Figure 7b, 90% (54/60) of the planets detected so far occur in the region where the planets can be explained by the core-accretion model. Furthermore, half of the planets detected occur in the region which could have been formed by core accretion in disks of MMSN ($f_g = 1$); $M_p < 3M_J$, $[Fe/H] > -0.11$. In addition, in the case of F-type and K-type dwarfs, as shown in Figure 7a and c, 80% of the detected planets occur in the region which could have been formed by core accretion. Moreover, in the case of M-type dwarfs (Figure 7d), although the number of samples is scarce, the trend is almost the same. Therefore, the hypothesis that almost all of the planets detected so far were formed by core accretion is supported by these observational facts.

To further examine the above hypothesis, we compare the observational facts with the theoretical identification in terms of the probability of gas giant formation due to metallicity of the central stars. According to the disk instability model, the probability of gas giant formation is independent of the metallicity of the host stars, [Fe/H]. On the other hand, according to the core-accretion model, there should be a lower limit for disk metallicity for gas giant formation. As this lower limit is inversely proportional to the total mass of a disk, it should be possible to determine which model is correct, when both total mass and metallicity of the disk are measured. However, it is impossible to find the total mass of the disk at the time of planetary formation. Therefore, assuming that the number distribution of the disk mass in the Taurus region ($[Fe/H] \sim 0$) observed by Andrews & Williams (2005) is approved in the overall region of metallicity, the percentage of gas giant formation is theoretically derived. In this regard, according to the MMSN model, we assume that the index of the gas surface density distribution for the semi-major axis is $-3/2$, and the outer radius of the disks is 250 AU (see section 5.3).

The fraction of disks where gas giants can be formed by core accretion with respect to all disks ($F(Z_{disk})$) is

$$F\left(Z_{disk}\right)=\frac{\int_{f_{g,crit}}^{\infty} N_D\left(f_g\right)df_g}{\int_{0}^{\infty} N_D\left(f_g\right)df_g}, \tag{21}$$

where $N_d\left(f_g\right)$ is the number distribution of the disk mass. In addition, according to equation (11), the lower mass of the disks for gas giant formation ($f_{g,cirt}$) is given by,

$$f_{g,cirt}^{\frac{17}{16}}\cdot 10^{Z_{disk}+0.11}\approx\left(\frac{\eta_{ice}}{4.2}\right)^{-1}\left(\frac{\tau_{disk}}{10^7\,yrs}\right)^{-\frac{5}{32}}\left(\frac{M_{c,crit}}{4M_{\oplus}}\right)^{\frac{59}{96}}\left(\frac{M_*}{M_{sun}}\right)^{\frac{49}{192}}. \tag{22}$$

Figure 8 shows the comparison of the percentage of gas giant formation expected from the core-accretion model with the fraction of detected gas giants from Fischer & Valenti (2005). The planets detected by Fischer & Valenti (2005) have a period of less than 4 years. In other words, supposing that all planets migrate to the orbits which have a period of less than 4 years, the observation would be directly compared with the theoretical results. As shown in Figure 8, although the percentage of gas giant formation expected from the core-accretion model is slightly higher than the observation, the trend is almost the same, and their values also fall within the error range. On the other hand, according to the disk instability model, the percentage gas giant formation should be constant. Therefore, it is likely that almost all of the planets detected so far are not formed by disk instability, but by core accretion.

### 5.2. *The Facts That Support The Disk Instability Scenario*

As described in section 5.1, in the case of G-type dwarfs, 90% (54/60) of the planets detected so far occur in the region where the gas giants can be explained by the core-accretion model. On the other hand, the other 10% (hereafter, massive gas giants) lie outside the region. The range where gas giants are thought to be formed through the core-accretion mechanism (shown in Figure 7b) is sensitive to the critical core mass $M_{c,crit}$ and the gas depletion timescale of the disks. Therefore, it is necessary to examine the validity of these adopted values, $M_{c,crit} = 4M_{\oplus}$ and $\tau_{disk} = 10^7 yrs$.

First, we discuss the gas depletion timescale of the disks. Haisch et al. (2001) and Wyatt et al. (2003) revealed that the gas depletion timescale of the disks ranges from 1 Myr to 10 Myrs based on YSO of submillimeter observation. Furthermore, Pascucci et al. (2006) found that the total gas mass contained within 40 AU of 8 YSO systems younger than ~30 Myr is less than $0.04 M_J$. These results indicate that the disks older than 10 Myr would not have sufficient gas for the formation of gas giants. Therefore, as assumed in section 2.1, the upper limit of the gas depletion timescale (10 Myr) is reasonable.

Next, we discuss the critical core mass. Ikoma, Emori, & Nakazawa (2000) estimated the critical core mass as

$$M_{c,crit} \sim 7\left(\frac{\dot{M}}{10^{-7} M_{\oplus} \cdot yr^{-1}}\right)^{0.2-0.3}\left(\frac{\kappa}{1cm^2 \cdot g^{-1}}\right)^{0.2-0.3} M_{\oplus}, \quad (23)$$

where $\dot{M}$ is the core's rate of planetesimal accretion and $\kappa$ is the opacity. Assuming that the gas depletion timescale of the disks is given as 10 Myr and the core's rate of planetesimal accretion is constant over time, $\dot{M} \sim M_{c,crit}/10^7$, the critical core mass is estimated as $M_{c,crit} \sim 13M_{\oplus}$ using equation (17). Ida & Lin (2004) indicated that, if the core's rate of planetesimal accretion is significantly reduced, $\dot{M} \sim 10^{-8} M_{\oplus}/yr$, the critical core mass may be $4M_{\oplus}$. However, when $M_c < 4M_{\oplus}$, gas accretion onto the core would not occur, because the Kelvin-Helmholtz timescale is longer than the gas depletion timescale of the disks. Therefore, $4M_{\oplus}$ is strictly a lower limit for the critical core mass.

On the other hand, as shown in Figure 7b, according to the disk instability model, the six detected massive gas giants formed by disk instability are 15 times the MMSN model ($f_g = 15$). In addition, although the sample number is scarce, the massive gas giants tend to be detected uninvolved with metallicity of the stars. Furthermore, in the case of F-type and K-type dwarfs, all detected massive gas giants can be formed by disk instability from disks which are 15 times the MMSN model ($f_g = 15$). As discussed in the next section, it is accepted that the massive disks, $f_g > 10$, exist at the time of disk formation. Conclusively, the massive gas giants of the detected planets are thought to be formed by disk instability.

### 5.3. *Gas Surface Density and Total Disk Mass*

In both the core-accretion and the disk instability model, the conditions for gas giant formation depend very much on the gas surface density of the disks. Therefore, we compare the gas surface density of the disks applied to the models with the observational data to validate whether that is reasonable. However, the timescale of disk instability is so small that in the case of very massive disks, the disk instability would not be detected.

First, we reveal the connection between the gas surface density of the disks and the total disk mass. For example, even if the MMSN model ($f_g = 1$) is accepted, the total mass of the disks is not determined uniquely. Kusaka et al. (1970) and Weidenschilling (1977) assume that all matter in the disks are distributed in the region of solar system formation, and estimate the total mass of the disks as $0.04 M_{sun}$ and $0.01 - 0.07 M_{sun}$, respectively. Kokubo & Ida (2002) and Ida & Lin (2004) adopt 0.013 $M_{\rm sun}$ for the total disk mass, which is adopted in this paper.

On the other hand, using high resolution imaging in the submillimeter wavelength, Mannings & Sargent (2000) and Kitamura et al. (2002) found that the radius of the disks in the Taurus region widen from 100 AU to 500 AU. (In section 5.1, these median values are adopted.) In addition, Kitamura et al. (2002) reveal that the gas surface density at 100 AU is consistent with that extrapolated for MMSN at 100 AU. Based on these observations, assuming that the gas surface density, similar to the MMSN model, is on the outer side of 36 AU, the total mass of the disks, $M_{disk}$, is given by

$$M_{disk} \sim \int_{a_{in}}^{a_{out}} da\Sigma(a)\cdot 2\pi a$$
$$\sim 4\pi a_{out}^2 \Sigma(a = a_{out})\left[1-\left(\frac{a_{in}}{a_{out}}\right)^{\frac{1}{2}}\right] \quad (24)$$

where $a_{in}$ is the inner radius of the disks and $a_{out}$ is the outer radius of the disks. In the case of $a_{out} = 100\,AU$ and $a_{out} = 500\,AU$, the total mass of the disks is about 0.03 $M_{\rm sun}$ and 0.07 $M_{\rm sun}$, respectively.

Andrews & Williams (2005) derived the number distribution of the disk mass in the Taurus region at submillimeter observation. In their results, the total mass of the disks ranges from 0.0005 $M_{\rm sun}$ to 0.5 $M_{\rm sun}$ for the 78 disks detected significantly. In the case where the outer radius is 100 AU and 500 AU, the ratio of the gas surface density to the MMSN model ranges from 0.01 to 17 and 0.005 to 7, respectively. Therefore, very massive disks that are more than 10 times the MMSN model ($f_{\rm g}$ = 10) do not exist. Submillimeter flux, however, does not reflect the mass of optically thick medium, and the real mass of the disks may be more massive than that observed.

Conclusively, it is reasonable to adopt the disks with $f_g > 10$ in the disk instability model; on the other hand, disks with $f_g < 10$ are adopted in the core-accretion model.

## 5.4. *Selection Effects*

We consider that the selection effects from the Doppler shift observation affect the above discussion. The radial velocity observation is sensitive to the massive planets which have a short period. Because the current detection limit is about 3 m/s, planets of 0.033 $M_{\rm J}$ at 0.1 AU and 0.106 $M_{\rm J}$ at 1 AU can be detected. In this paper, only planets of more than 0.1 $M_{\rm J}$ are selected as samples.

Measurement accuracy for F-type dwarfs is relatively poor, because these stars have less absorption lines and higher rotational velocity, compared with the G-type and K-type stars. Tinney et al. (2001), Fischer & Valenti (2003), and Johnson et al. (2006) show that the above effects operate and make the detection limit useless for earlier F7-type stars. In this paper, only star types later than F7 are selected as samples.

Finally, we discuss the effects of selection on metallicity of the central stars. The observational targets are biased for the metal-rich stars, because the frequency of detection of the planets is empirically high among these stars. Therefore, for the results where the detection number around the metal-poor stars is low, it is possible that these results have been affected by the selection effect. Therefore, in the discussion of metallicity dependence, the results derived by Fischer & Valenti (2005) are adopted.

## 6. CONCLUSION

We derived the conditions for metallicity of the disks and planet mass for gas giant formation using the core-accretion and the disk instability model. Next, we checked whether the planets detected (161 cases) so far satisfy the above conditions for each spectral type of the central stars. As a result, regardless of the spectral type, 90% of the planets detected occur in the range where gas giants can be formed by core accretion. On the other hand, about another 10% are in the range where gas giants can be explained by the disk instability model, but not by the core-accretion model. Next, we derived the metallicity dependence for the percentage of gas giant formation in the case of the core-accretion model. Comparing it with the fraction of gas giants detected ($M_p$ > 0.1 $M_J$), both sides are found to be consistent. On the other hand, the observation cannot be explained by the disk instability model, because the condition for gas giant formation is independent of metallicity. Therefore, most of planets detected so far are thought to have been formed by core-accretion, and the rest by the disk instability process.

We thank S. Ida, S. Watanabe, and H. Tanaka for useful discussions and comments. This study is supported by a Grant-in-Aid provided from the Japan Society of Promotion of Science (JSPS).

## REFERENCES

Alibert, Y., Mordasini, C., & Benz, W. 2004, A&A, 417, L25
Alibert, Y., Mordasini, C., Benz, W., & Winisdoerffer, C. 2005, A&A, 434, 343
Andrews, S. M., & Williams, J. P. 2005, ApJ, 631, 1134

Bachiller R. 1996, ARA&A, 34, 111
Beckwith S. V. W., & Sargent A. I., 1996, Nature, 383, 139
Beckwith S. V. W., Sargent A. I., & Chini R. S., 1990, AJ, 99, 924
Bonfils X., et al. 2005, A&A, 443, L15
Boss, A. P. 1997, Science, 276, 1836
Boss, A. P. 2002, ApJ, 567, L149
Cabot, W., Canuto, V. M., Hubickyj, O., & Pollack J. B. 1987, Icarus, 69, 423
Cai, K., et al. 2006, ApJ, 636, L149
Cameron, A. G. W. 1978, Moon Planets 18, 5
D'Anglelo G., Henning T., & Kley W., 2002, A&A, 385, 647
D'Anglelo G., Kley W., & Henning T., 2003, A&A, 586, 540
Endl, M., et al. 2004, ApJ, 611. 1121
Fischer, D. A., & Valenti, J. A. 2003. in ASP Conf. Ser. 294, Scientific Frontiers in Research on Extrasolar Planets, ed. D. Deming & S. Seager. (SanFrancisco: ASP) , 117
Fischer, D. A.& Valenti, J. A. 2005, ApJ, 622, 1102
Fukagawa, M., et al. 2004, ApJ. 605, L53
Gammie C. F. 2001, ApJ. 553, 174
Goldreich, P. & Ward, W. 1973, ApJ, 183, 1051
Gonzalez, G. 1997, MNRAS 285, 403
Haisch K. E. Jr., Lada E. A., & Lada C. J. 2001, ApJ, 553, L153
Hayashi, C. 1981, Prog. Theor. Phys. Suppl., 70, 35
Hayashi, C., Nakazawa, K., & Nakagawa Y. 1985, in Protostars and Planets Ⅱ,ed. D. C. Black & M. S. Matthew (Tueson: Univ. Arizona Press), 1100
Ida, S.& Lin D. N. C. 2004a, ApJ, 604, 388
Ida, S., & Lin, D. N. C. 2004b, ApJ, 616, 567
Ikoma, M., Nakazawa, K., & Emori, H. 2000, ApJ, 537, 1013
Johnson, B. M., & Gammie, C. F. 2003, ApJ, 597, 131
Kitamura, Y., et al. 2002, ApJ, 581, 357
Kokubo, E., & Ida, S. 2002, ApJ, 581, 666
Kuiper, G. P. 1951, Proc. Natl. Acad. Sci. U.S.A., 37, 1
Kusaka, T., Nakano, T., & Hayashi, C., 1970, Prog. Theor. Phys. Suppl., 44, 1580
Laughlin, G. & Bodenheimer, P. 1994, ApJ, 436, 335
Laws, C., et al. 2003, AJ, 125, 2664
Lovis, C., et al. 2006, Nature, 441, 305

Mejía, A. C., Durisen, R. H., Pickett, M. K., & Cai K., 2005, ApJ, 619, 1098
Mayer, L., et al. 2002, Science, 298, 1756
Mayer, P., Quinn, T., Wadsley, J., & Stadel, J. 2004, ApJ, 609, 1045
Nordstrom, R., et al. 2004, A&A, 419, 989
Pickett, et al. 2003, ApJ, 590, 1060
Pollack, J. B., et al. 1996, Icarus, 124, 62
Rice, W. K. M., Armitage, P. J., & Bonnell, I. A. 2003, MNRAS, 346, 36
Robinson, E. S., Laughlin, G., Bodenheimer., & Fischer, D. 2006, ApJ, 643, 484
Safronov, V. 1969, Evolution of the Protoplanetary Cloud and Formation of the Earth and Planets (Moscow: Nauka)
Santos, N. C., Israelian, G., & Mayor, M. 2004, A&A, 415, 1153
Santos, N. C., et al. 2005, A&A, 437, 1127
Sato, B., et al. 2005, ApJ, 633, 465
Saumon, D., & Guillot, T. 2003, American Astronomical Society, DPS meeting #35, #45.01, Bulletin of the American Astronomical Society, 35, 1008 35, 4501
Schuler, S. C., et al. 2005, ApJ, 632, L131
Setiawan, J., et al. 2005, A&A, 437, L31
Da Silva, R., et al. 2005, A&A, 446, 717
Toomre, A. 1964, ApJ, 139, 1217
Udry, S., et al. 2006, A&A, 447, 361
Valenti, J. A., & Fischer, D. A., 2005, ApJS, 159, 141
Weidenschilling, S. J. 1977, Astrophysics and Space Science, 51, 153
Wyatt, M. C., Dent, W. R. F., & Greaves, J. S. 2003, MNRAS, 342, 876
Young, R. E. 2003, New Astronomy Reviews, 47, 1

Table1

| St. Name | St. [Fe/H] | St. Spec. Type | Reference |
|---|---|---|---|
| 14 Her | 0.43±0.08 | K0 V | Santos et al. 2004 |
| 16 Cyg B | 0.08±0.04 | G2.5 V | Santos et al. 2004 |
| 47 Uma | 0.05±0.02<br>0.06±0.03<br>0.05±0.02 | G0V | Laws et al. 2003<br>Santos et al. 2004<br>Average |
| 51 Peg | 0.2±0.05 | G2 IV | Santos et al. 2004 |
| 55 Cnc | 0.33±0.07 | G8 V | Santos et al. 2004 |
| 70 Vir | -0.02±0.04<br>-0.06±0.05<br>-0.04±0.03 | G4 V | Laws et al. 2003<br>Santos et al. 2004<br>Average |
| BD-10 3166 | 0.35±0.05 | G4 V | Santos et al. 2005 |
| Epsilon Eridani | -0.09±0.03<br>-0.13±0.04<br>-0.10±0.02 | K2 V | Laws et al. 2003<br>Santos et al. 2004<br>Average |
| Gamma Cephei | 0.16±0.08 | K2 V | Santos et al. 2004 |
| GJ 3021 | 0.12±0.06 | G6 V | Santos et al. 2005 |
| GJ 436 | -0.03 | M2.5 | Bonfile et al. 2005 |
| Gl 581 | -0.25 | M3 | Bonfile et al. 2005 |
| Gl 86 | -0.25±0.05<br>-0.23±0.04<br>-0.24±0.03 | K1V | Santos et al. 2004<br>Santos et al. 2004<br>Average |
| Gliese 876 | 0.02 | M4 V | Bonfile et al. 2005 |
| HD 101930 | 0.17±0.06 | K1 V | Santos et al. 2005 |
| HD 102117 | 0.33±0.06<br>0.30±0.03<br>0.31±0.03 | G6V | Santos et al. 2005<br>Santos et al. 2005<br>Average |
| HD 104985 | -0.28±0.09 | G9 III | Santos et al. 2005 |
| HD 106252 | -0.05±0.03<br>-0.01±0.05<br>-0.03±0.05<br>-0.04±0.02 | G0 | Laws et al. 2003<br>Santos et al. 2004<br>Santos et al. 2005<br>Average |
| HD 10647 | -0.03±0.04 | F8V | Santos et al. 2004 |
| HD 10697 | 0.14±0.04 | G5 IV | Santos et al. 2004 |
| HD 108147 | 0.23±0.06<br>0.20±0.05<br>0.21±0.04 | F8/G0 V | Laws et al. 2003<br>Santos et al. 2004<br>Average |
| HD 108874 | 0.23±0.05 | G5 | Santos et al. 2004 |
| HD 109749 | 0.25±0.05 | G3 IV | Fischer et al. 2005 |
| HD 111232 | -0.36±0.04 | G8V | Santos et al. 2004 |
| HD 114386 | -0.08±0.06<br>-0.04±0.07<br>-0.06±0.05 | K3 V | Santos et al. 2004<br>Santos et al. 2005<br>Average |
| HD 114729 | -0.25±0.05 | G3 V | Santos et al. 2004 |
| HD 114783 | 0.17±0.02<br>0.09±0.04<br>0.15±0.02 | K0 | Laws et al. 2003<br>Santos et al. 2004<br>Average |
| HD 117207 | 0.23±0.05 | G8VI/V | Santos et al. 2005 |
| HD 117618 | 0.06±0.06 | G2V | Santos et al. 2005 |
| HD 118203 | 0.1±0.05 | K0 | Da Silva et al. 2005 |
| HD 11977 | -0.21±0.10 | G8.5 III | Setiawan et al. 2005 |
| HD 121504 | 0.12±0.05<br>0.16±0.05<br>0.14±0.04 | G2 V | Laws et al. 2003<br>Santos et al. 2004<br>Average |
| HD 12661 | 0.36±0.05 | G6 V | Santos et al. 2004 |
| HD 128311 | 0.03±0.07 | K0 | Santos et al. 2004 |
| HD 130322 | 0.03±0.04 | K0 V | Santos et al. 2004 |
| HD 13189 | -0.58±0.04 | K2 II | Schuler et al. 2005 |
| HD 134987 | 0.3±0.04 | G5 V | Santos et al. 2004 |
| HD 136118 | -0.05±0.03<br>-0.04±0.05<br>-0.04±0.03 | F9 V | Laws et al. 2003<br>Santos et al. 2004<br>Average |
| HD 141937 | 0.14±0.04<br>0.10±0.05<br>0.12±0.03 | G2/G3 V | Laws et al. 2003<br>Santos et al. 2004<br>Average |
| HD 142 | 0.14±0.07 | G1 IV | Santos et al. 2004 |
| HD 142022 A | 0.19±0.04 | K0 V | Santos et al. 2005 |
| HD 142415 | 0.21±0.05 | G1 V | Santos et al. 2004 |
| HD 147513 | 0.08±0.04<br>0.06±0.04<br>0.07±0.03 | G3/G5V | Santos et al. 2004<br>Santos et al. 2005<br>Average |
| HD 149026 | 0.36±0.05 | G0 IV | Sato et al. 2005 |
| HD 149143 | 0.26±0.05 | G0 IV | Fischer et al. 2005 |
| HD 150706 | -0.01±0.04 | G0 | Santos et al. 2004 |
| HD 154857 | -0.23±0.04 | G5V | Santos et al. 2005 |
| HD 160691 | 0.28±0.03<br>0.32±0.04<br>0.32±0.05<br>0.3±0.02 | G3 IV-V | Laws et al. 2003<br>Santos et al. 2004<br>Santos et al. 2005<br>Average |
| HD 16141 | 0.19±0.03<br>0.15±0.04<br>0.18±0.02 | G5 IV | Laws et al. 2003<br>Santos et al. 2004<br>Average |
| HD 162020 | -0.09±0.07<br>-0.01±0.08<br>-0.06±0.05 | K2 V | Santos et al. 2004<br>Santos et al. 2004<br>Average |
| HD 168443 | 0.06±0.05 | G5 | Santos et al. 2004 |
| HD 168746 | -0.06±0.03<br>-0.08±0.05<br>-0.07±0.03 | G5 | Laws et al. 2003<br>Santos et al. 2004<br>Average |
| HD 169830 | 0.17±0.04<br>0.21±0.05<br>0.19±0.03 | F8 V | Laws et al. 2003<br>Santos et al. 2004<br>Average |
| HD 177830 | 0.33±0.09 | K0 | Santos et al. 2004 |
| HD 178911 B | 0.24±0.10<br>0.27±0.05<br>0.26±0.04 | G5 | Santos et al. 2004<br>Santos et al. 2004<br>Average |
| HD 179949 | 0.22±0.05 | F8 V | Santos et al. 2004 |
| HD 183263 | 0.34±0.04 | G2IV | Santos et al. 2005 |
| HD 187123 | 0.13±0.03 | G5 | Santos et al. 2004 |
| HD 188015 | 0.3±0.05 | G5IV | Santos et al. 2005 |
| HD 190228 | -0.24±0.03<br>-0.25±0.05<br>-0.27±0.06<br>-0.25±0.02 | G5IV | Laws et al. 2003<br>Santos et al. 2004<br>Santos et al. 2004<br>Average |
| HD 190360 | 0.24±0.05 | G6 IV | Santos et al. 2004 |
| HD 192263 | -0.02±0.06 | K2 V | Santos et al. 2004 |
| HD 195019 | 0.03±0.03<br>0.09±0.04<br>0.06±0.05<br>0.05±0.02 | G3 IV-V | Laws et al. 2003<br>Santos et al. 2004<br>Santos et al. 2004<br>Average |
| HD 196050 | 0.22±0.05 | G3 V | Santos et al. 2004 |
| HD 19994 | 0.14±0.04<br>0.25±0.08<br>0.32±0.07<br>0.19±0.05<br>0.21±0.08<br>0.20±0.03 | F8 V | Laws et al. 2003<br>Santos et al. 2004<br>Santos et al. 2004<br>Santos et al. 2004<br>Santos et al. 2004<br>Average |
| HD 202206 | 0.33±0.03<br>0.35±0.06<br>0.33±0.03 | G6 V | Laws et al. 2003<br>Santos et al. 2004<br>Average |
| HD 20367 | 0.17±0.10 | G0 | Santos et al. 2004 |
| HD 2039 | 0.32±0.06 | G2/G3 IV-V | Santos et al. 2004 |
| HD 208487 | 0.06±0.04 | G2V | Santos et al. 2005 |
| HD 209458 | 0.02±0.03 | G0 V | Santos et al. 2004 |
| HD 210277 | 0.21±0.04<br>0.16±0.04<br>0.19±0.03 | G0 | Santos et al. 2004<br>Santos et al. 2004<br>Average |
| HD 213240 | 0.23±0.06<br>0.17±0.05<br>0.19±0.04 | G4 IV | Laws et al. 2003<br>Santos et al. 2004<br>Average |
| HD 216435 | 0.22±0.05 | G0 V | Santos et al. 2004 |
| HD 216437 | 0.25±0.04 | G4 IV-V | Santos et al. 2004 |
| HD 216770 | 0.26±0.04 | K1 V | Santos et al. 2004 |
| HD 217107 | 0.37±0.05 | G8 IV | Santos et al. 2004 |
| HD 219449 | 0.05±0.14 | K0 III | Santos et al. 2005 |
| HD 222582 | 0.05±0.05 | G5 | Santos et al. 2004 |
| HD 23079 | -0.11±0.06 | F8/G0 V | Santos et al. 2004 |
| HD 2638 | 0.16±0.04 | G5 | Santos et al. 2005 |
| HD 27442 | 0.41±0.05 | K2 IV a | Laws et al. 2003 |
| | 0.39±0.13<br>0.41±0.05 | | Santos et al. 2004<br>Average |
| HD 27894 | 0.30±0.07 | K2 V | Santos et al. 2005 |
| HD 28185 | 0.24±0.02<br>0.22±0.05<br>0.24±0.02 | G5 | Laws et al. 2003<br>Santos etal. 2004<br>Average |
| HD 30177 | 0.39±0.06<br>0.38±0.09<br>0.39±0.05 | G8 V | Santos etal. 2004<br>Santos etal. 2004<br>Average |
| HD 330075 | 0.08±0.06 | G5 | Santos et al. 2005 |
| HD 33564 | -0.12 | F6 V | Nordstrom et al. 2004 |
| HD 33636 | -0.11±0.03<br>-0.08±0.06<br>-0.104±0.03 | G0 V | Laws et al. 2003<br>Santos et al. 2005<br>Average |
| HD 3651 | 0.12+0.04 | K0 V | Santos et al. 2004 |
| HD 37124 | -0.37±0.03<br>-0.38±0.04<br>-0.37±0.02 | G4 V | Laws et al. 2003<br>Santos et al. 2004<br>Average |
| HD 37605 | 0.31±0.06 | K0V | Santos et al. 2005 |
| HD 38529 | 0.40±0.06 | G4 IV | Santos et al. 2004 |
| HD 39091 | 0.10±0.04 | G1 IV | Santos et al. 2004 |
| HD 40979 | 0.21±0.05 | F8 V | Santos et al. 2004 |
| HD 41004 A | 0.16±0.07 | K1 V | Santos et al. 2005 |
| HD 4203 | 0.40±0.04<br>0.40±0.05<br>0.40±0.03 | G5 | Laws et al. 2003<br>Santos et al. 2004<br>Average |
| HD 4208 | -0.25±0.03<br>-0.24±0.04<br>-0.25±0.02 | G5 V | Laws et al. 2003<br>Santos et al. 2004<br>Average |
| HD 4308 | -0.31±0.01 | G5 V | Udry et al. 2006 |
| HD 45350 | 0.29 | G5 IV | Valenti & Fischer 2005 |
| HD 46375 | 0.30±0.03<br>0.20 ±0.06<br>0.28±0.03 | K1 IV | Laws et al. 2003<br>Santos et al. 2004<br>Average |
| HD 47536 | -0.54±0.12 | K1 III | Santos et al. 2004 |
| HD 49674 | 0.33±0.06 | G5 V | Santos et al. 2004 |
| HD 50554 | 0.02±0.02<br>0.01±0.04<br>0.02±0.02 | F8 | Laws et al. 2003<br>Santos et al. 2004<br>Average |
| HD 52265 | 0.20±0.04<br>0.25±0.03<br>0.22±0.03 | G0 V | Santos et al. 2004<br>Santos et al. 2004<br>Average |
| HD 59686 | 0.28±0.18 | K2 III | Santos et al. 2005 |
| HD 63454 | 0.11 | K4 V | Santos et al. 2005 |
| HD 6434 | -0.55±0.07<br>-0.52±0.08<br>-0.54±0.05 | G3 IV | Laws et al. 2003<br>Santos et al. 2004<br>Average |
| HD 65216 | -0.12±0.04 | G5 V | Santos et al. 2004 |
| HD 68988 | 0.34±0.04<br>0.36±0.06<br>0.35±0.03 | G0 | Laws et al. 2003<br>Santos et al. 2004<br>Average |
| HD 69830 | -0.05±0.02 | K0V | Lovis et al. 2006 |
| HD 70642 | 0.18±0.04<br>0.20±0.06<br>0.19±0.03 | G5 IV-V | Santos et al. 2004<br>Santos et al. 2005<br>Average |
| HD 72659 | 0.03±0.06 | GO V | Santos et al. 2004 |
| HD 73256 | 0.29 | G8/K0 | Santos et al. 2004 |
| HD 73526 | 0.27±0.06 | G6 V | Santos et al. 2004 |
| HD 74156 | 0.16±0.05 | G0 | Santos et al. 2004 |
| HD 75289 | 0.28±0.07 | G0 V | Santos et al. 2004 |
| HD 76700 | 0.41±0.08 | G6 V | Santos et al. 2004 |
| HD 80606 | 0.32±0.09 | G5 | Santos et al. 2004 |
| HD 82943 | 0.26±0.03<br>0.32±0.05<br>0.29±0.02<br>0.28±0.02 | G0 | Laws et al. 2003<br>Santos et al. 2004<br>Santos et al. 2004<br>Average |
| HD 83443 | 0.36±0.04<br>0.35±0.08<br>0.39±0.07 | K0 V | Laws et al. 2003<br>Santos et al. 2004<br>Santos et al. 2004 |
| | 0.36±0.03 | | Average |
| HD 8574 | 0.02±0.03<br>0.06±0.07<br>0.03±0.03 | F8 | Laws et al. 2003<br>Santos et al. 2004<br>Average |
| HD 88133 | 0.33±0.05 | G5 IV | Santos et al. 2005 |
| HD 89744 | 0.22±0.05 | F7 V | Santos et al. 2004 |
| HD 92788 | 0.32±0.05<br>0.34±0.05<br>0.33±0.04 | G5 | Santos et al. 2004<br>Santos et al. 2005<br>Average |
| HD 93083 | 0.15±0.04 | K3 V | Santos et al. 2005 |
| HD 99492 | 0.26±0.07 | K2V | Santos et al. 2005 |
| HR 810 | 0.26±0.06 | G0V pecul. | Santos et al. 2004 |
| rho CrB | -0.21±0.04 | G0V or G2V | Santos et al. 2004 |
| Tau Boo | 0.23±0.07 | F7 V | Santos et al. 2004 |
| Ups And | 0.13±0.08 | F8 V | Santos et al. 2004 |

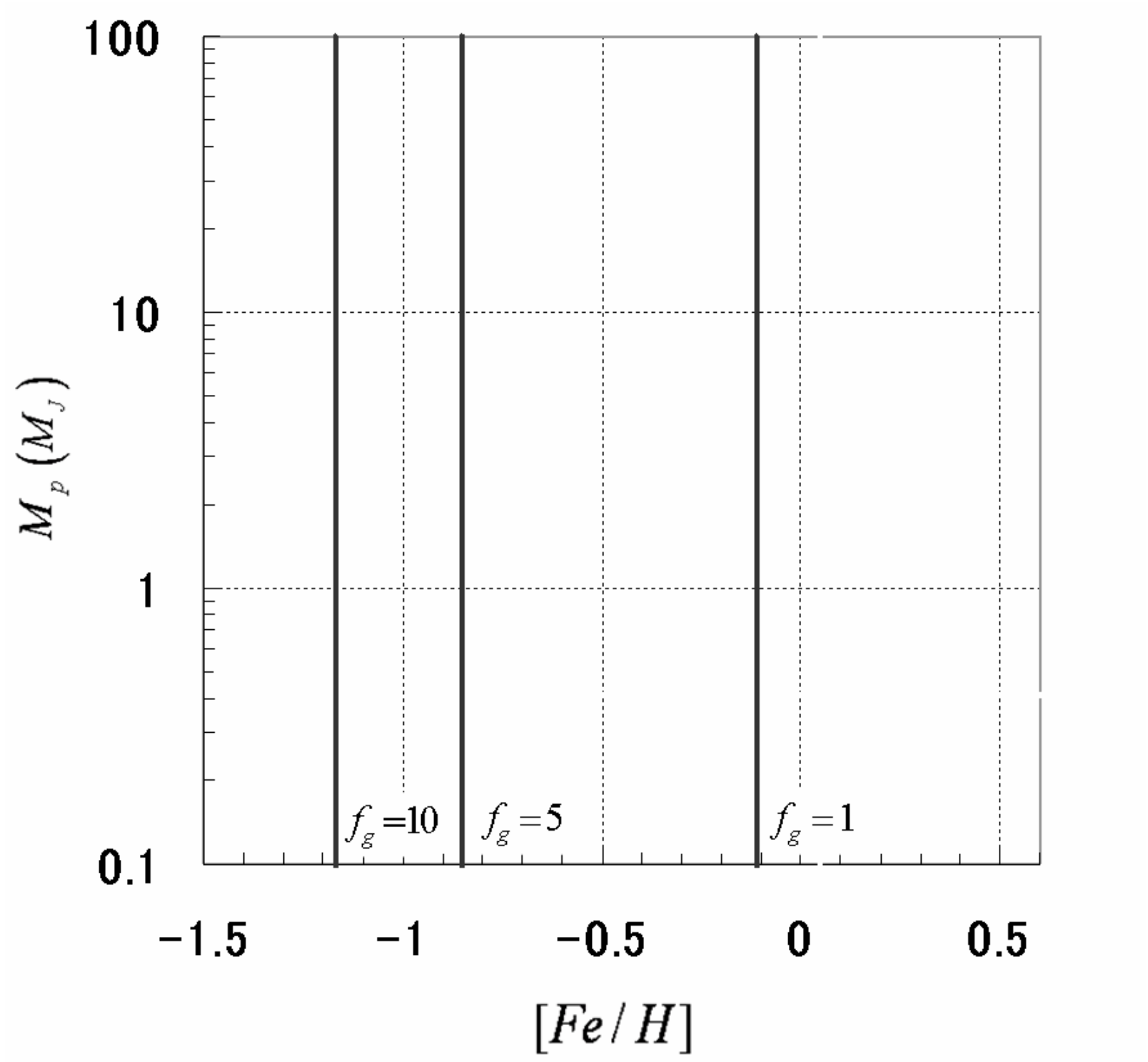

Figure 1: The metallicity condition for gas giants formed by core accretion in the case of G-type stars. The horizontal axis is metallicity of the disk and the vertical axis is the planet mass. The solid lines show the lower limits for gas giant formation. $f_g$ represents the scaling factor for the gas surface density of the disk.

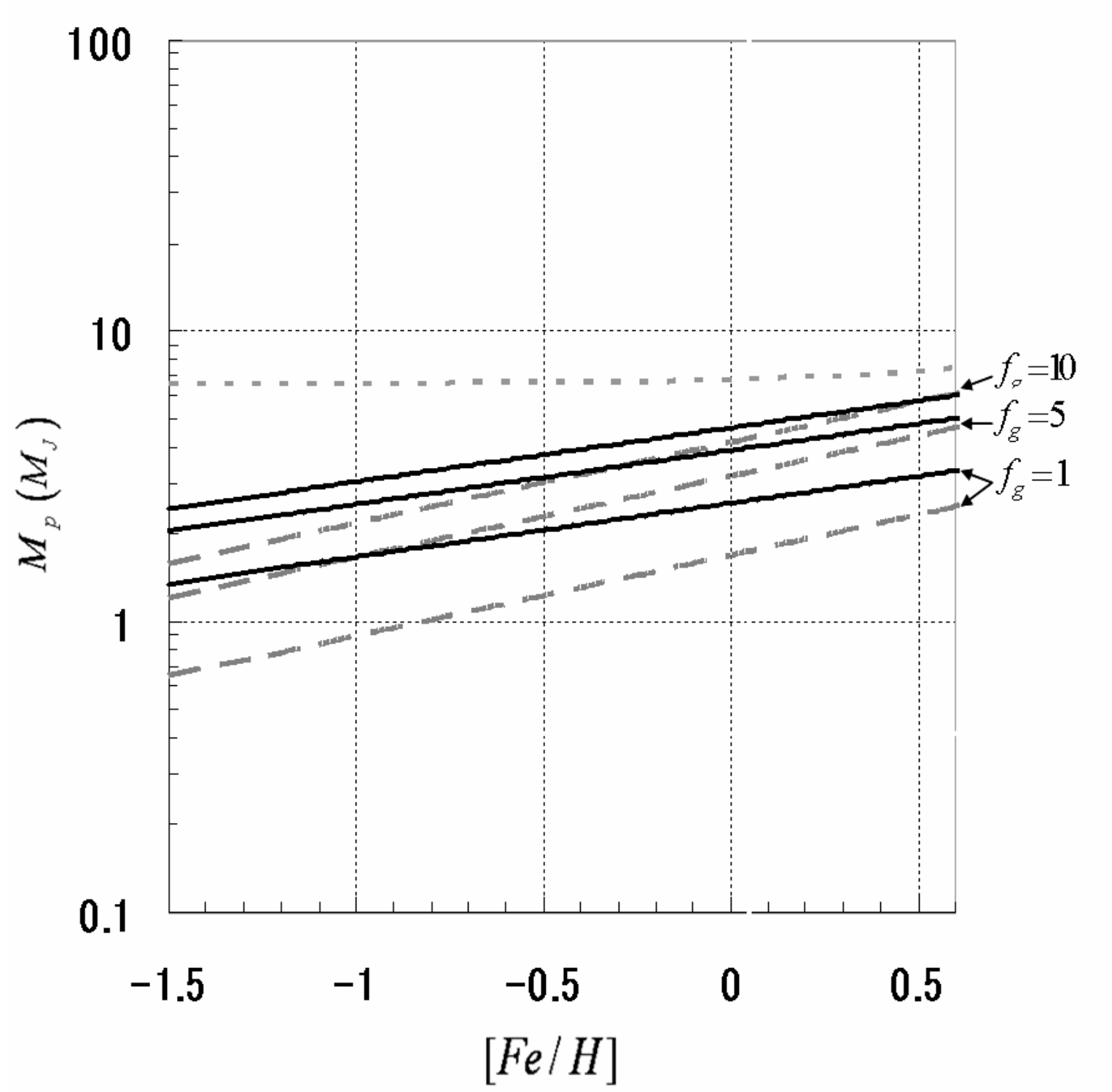

Figure 2: The upper limits for the mass of gas giant formed by core accretion in the case of G-type stars as a function of metallicity. The horizontal axis is metallicity of the disk and the vertical axis is the planet mass. The dashed and solid lines show $M_{p,\max 1}$ and $M_{p,\max 2}$, respectively. The dotted line shows the planet in which the gas around the gap accretes onto the planet with $M_{p,\max 1}$ over 10 Myrs, after a gap is formed. $f_g$ represents the scaling factor for the gas surface density of the disk.

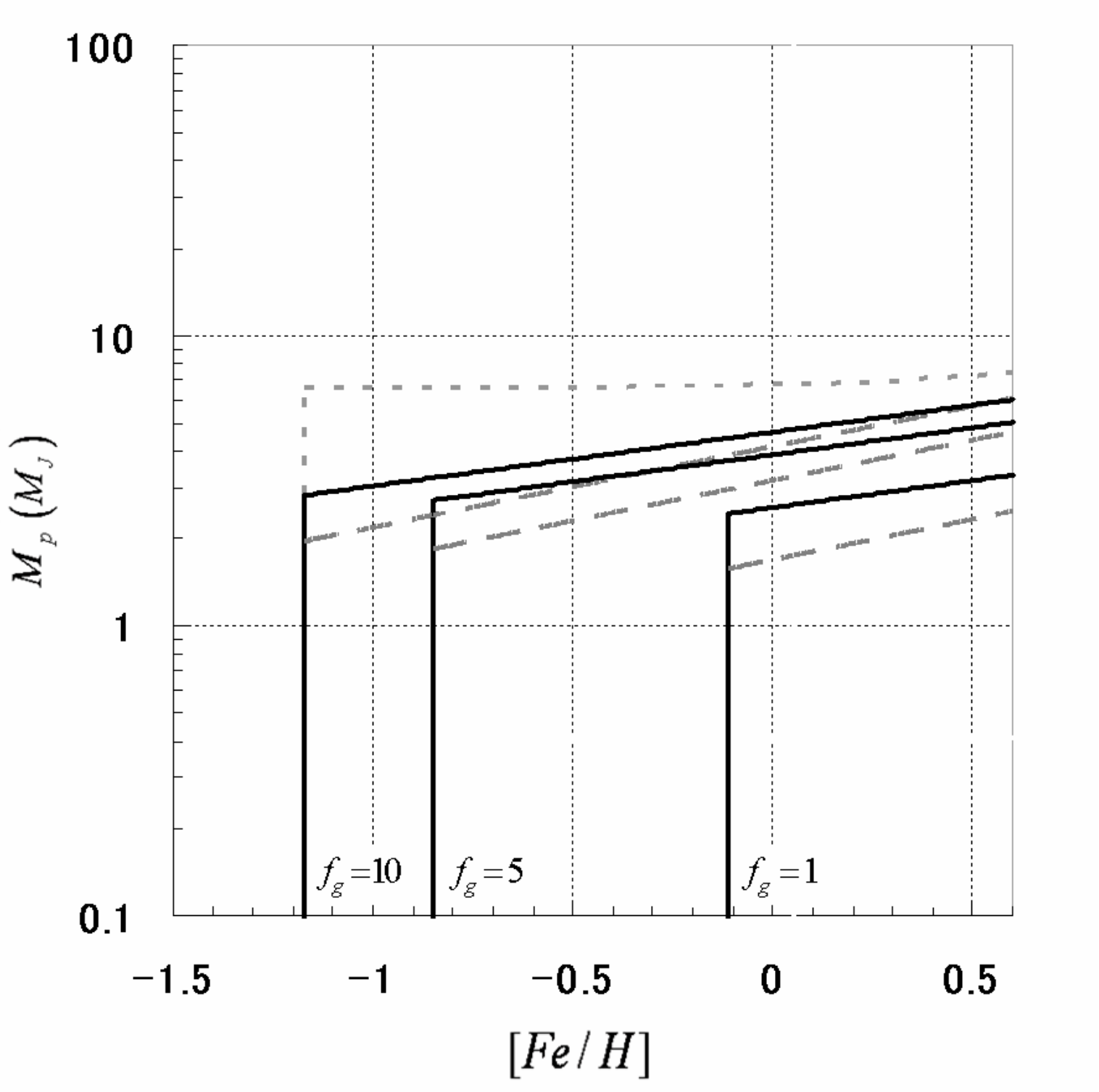

Figure 3: The condition of disk metallicity for gas giant formation through the core-accretion mechanism and the upper limits for planet mass in the case of G-type stars. The horizontal axis is metallicity of the disk and the vertical axis is the planet mass. The dotted line shows the planet in which the gas around the gap accretes onto the planet with $M_{p,\max 1}$ over 10 Myrs, after a gap is formed. $f_g$ represents the scaling factor for the gas surface density of the disk. The bottom and right-hand side of each boundary line is the allowed region for gas giant formation in the core-accretion model.

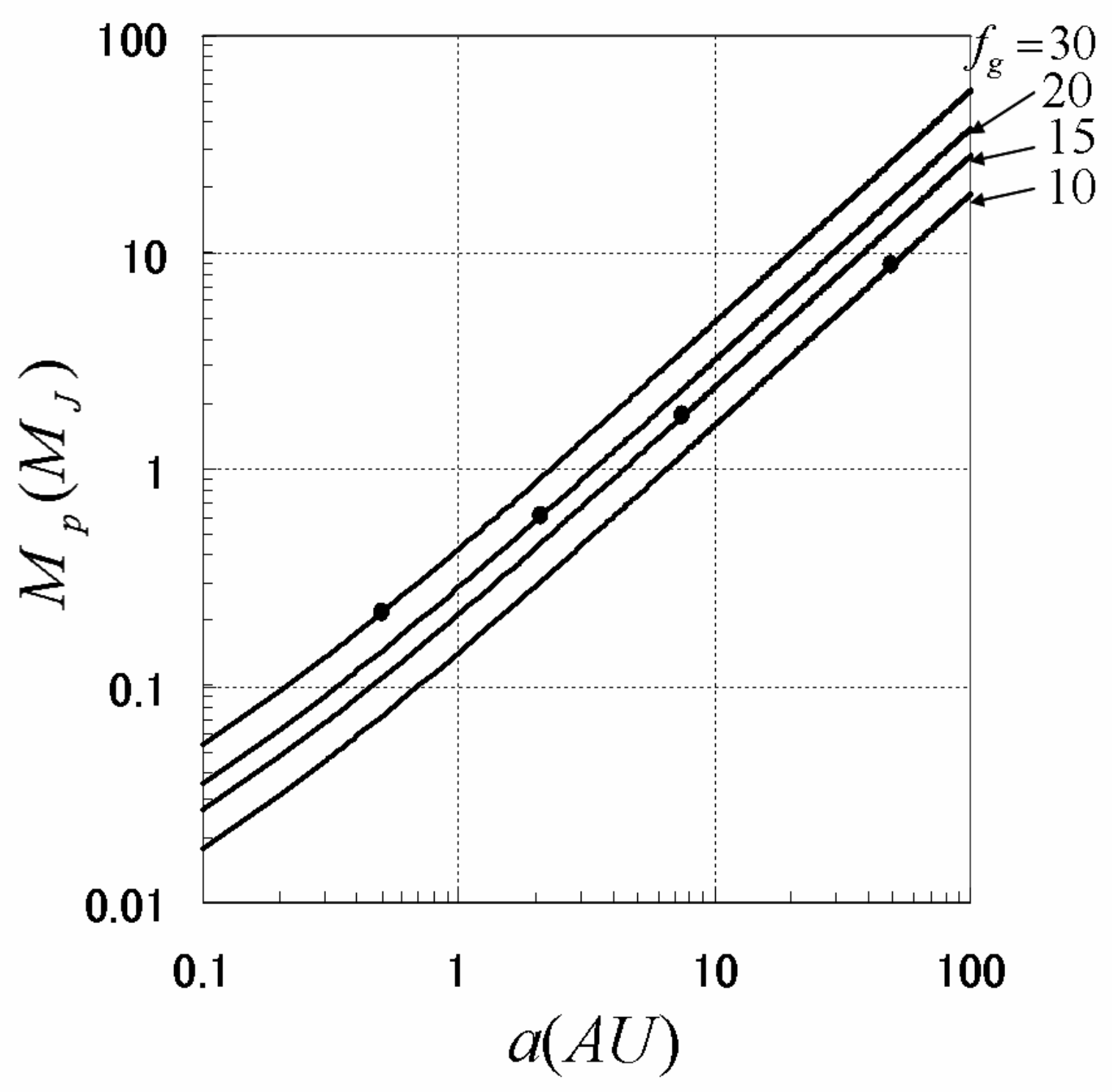

Figure 4: The lower limits for the mass of the gas giant formed by disk instability in the case of G-type stars as a function of the semi-major axis. The solid line represents the minimum mass of the planet formed by disk instability for the semi-major axis. The dots represent the semi-major axes where the Toomre-Q value is equal to 1 and the minimum mass of the planets is formed. $f_g$ represents the scaling factor for the gas surface density of the disk.

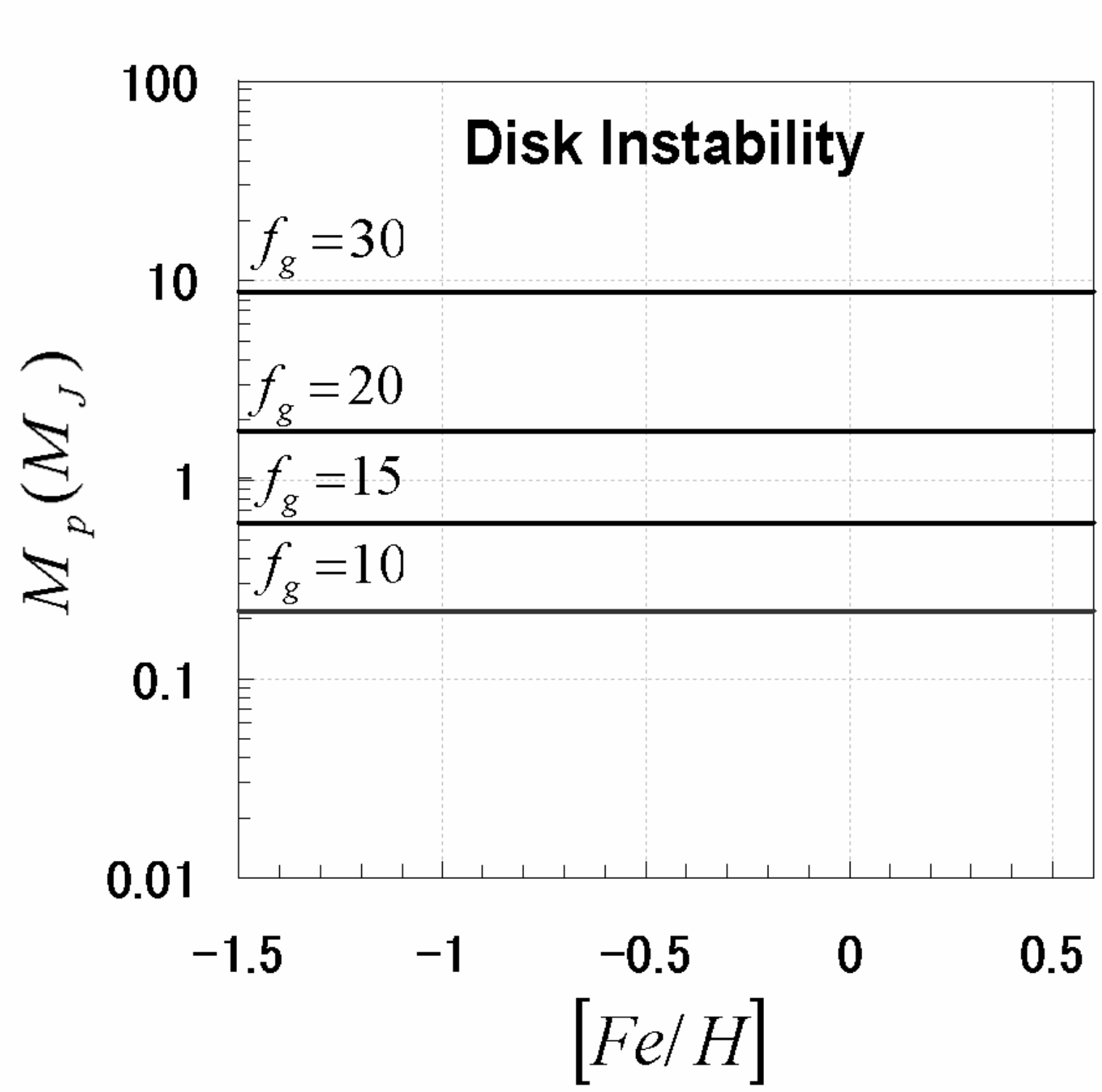

Figure 5: The condition for metallicity of the disks and the planet mass for gas giant formation by disk instability in the case of G-type stars. The horizontal and vertical axes represent metallicity of the disks and the planet mass, respectively. The solid lines show the lower limits of the planet mass. The region where gas giants can be formed by disk instability is the upper side of the boundary line. $f_g$ represents the scaling factor for the gas surface density of the disk.

Figure 6: The correlation between metallicity of the host stars and the planet mass

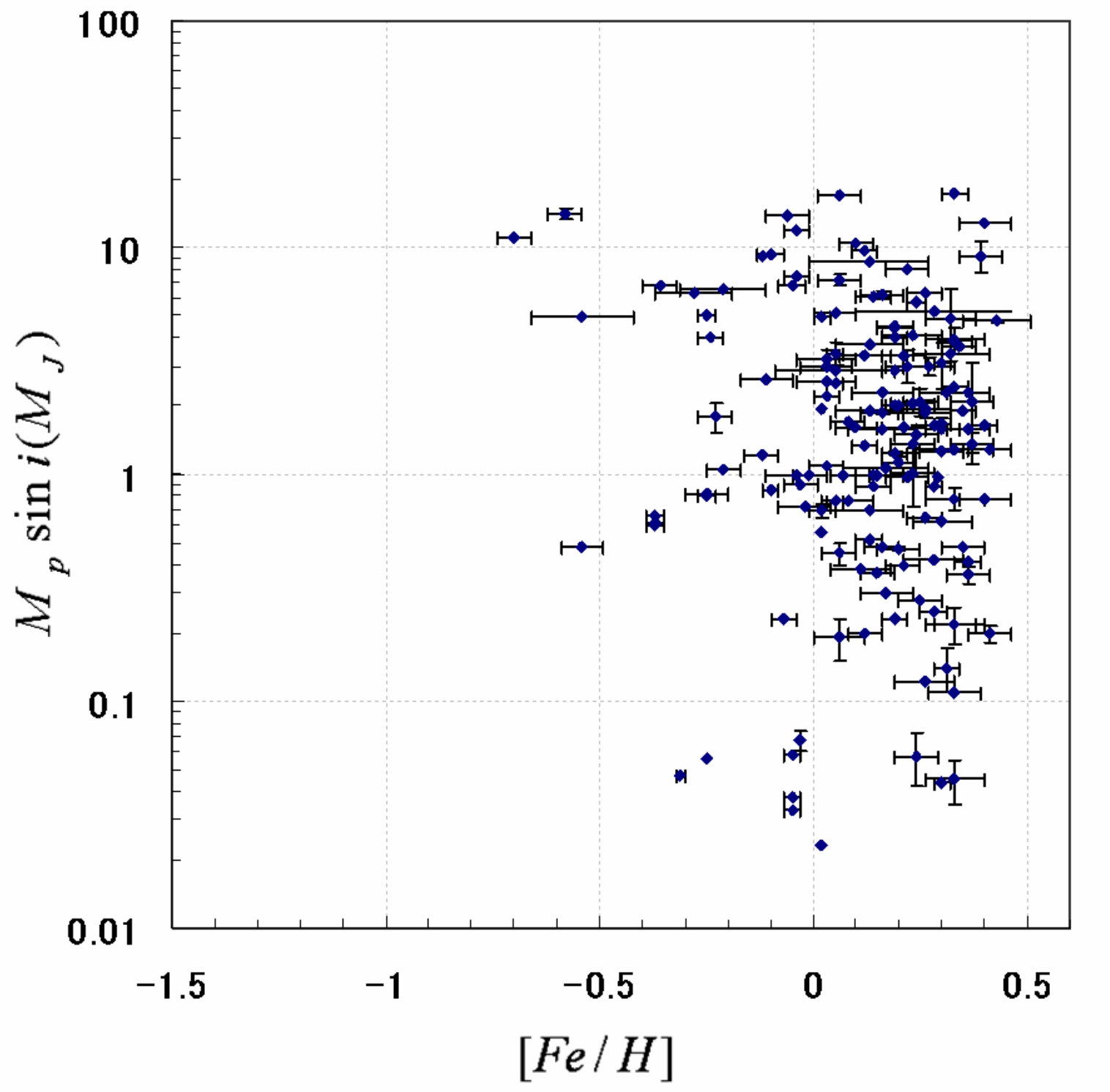

Figure 6a: For all samples. (161 samples)

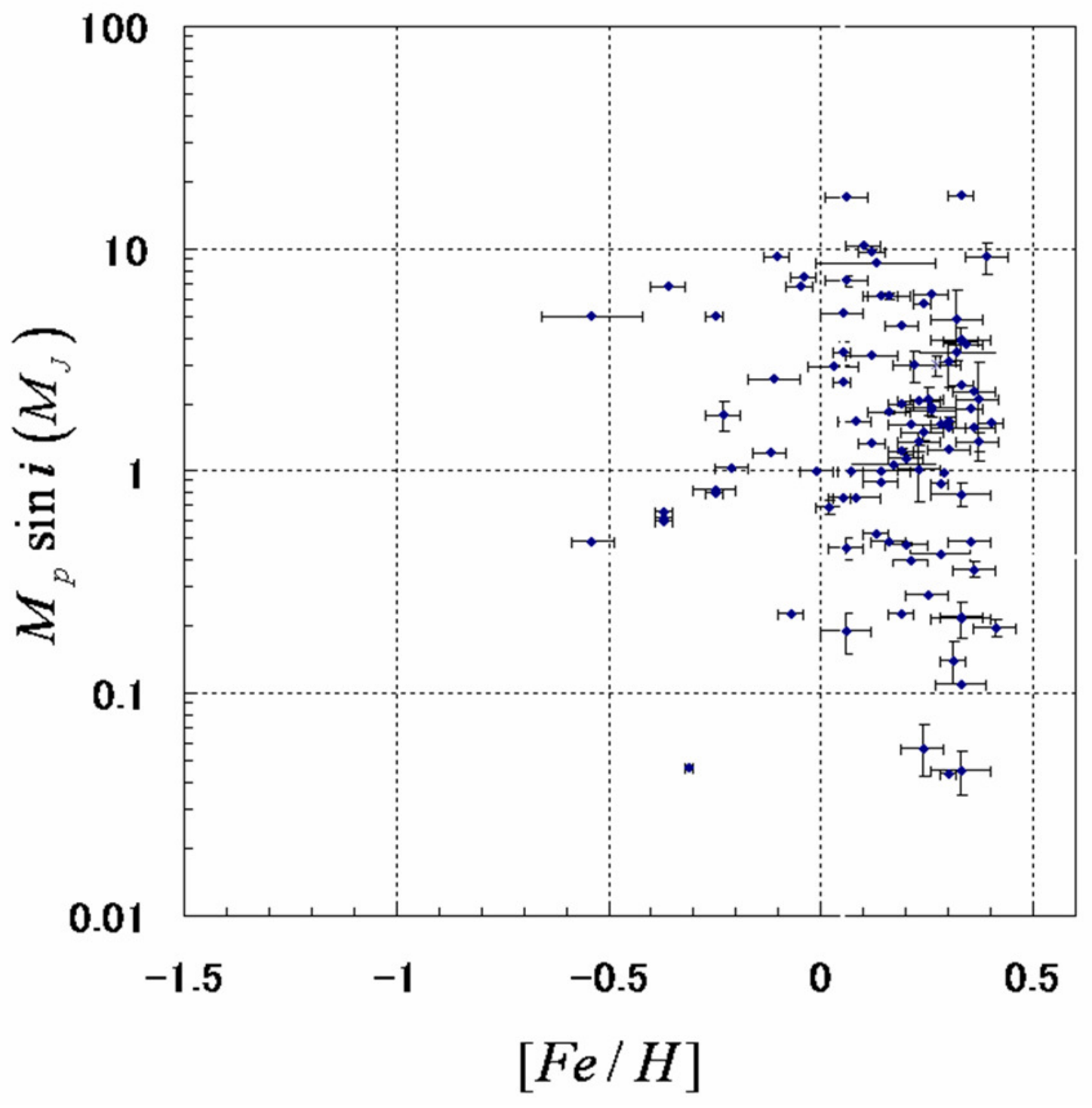

Figure 6b: For G-type stars. (105 samples)

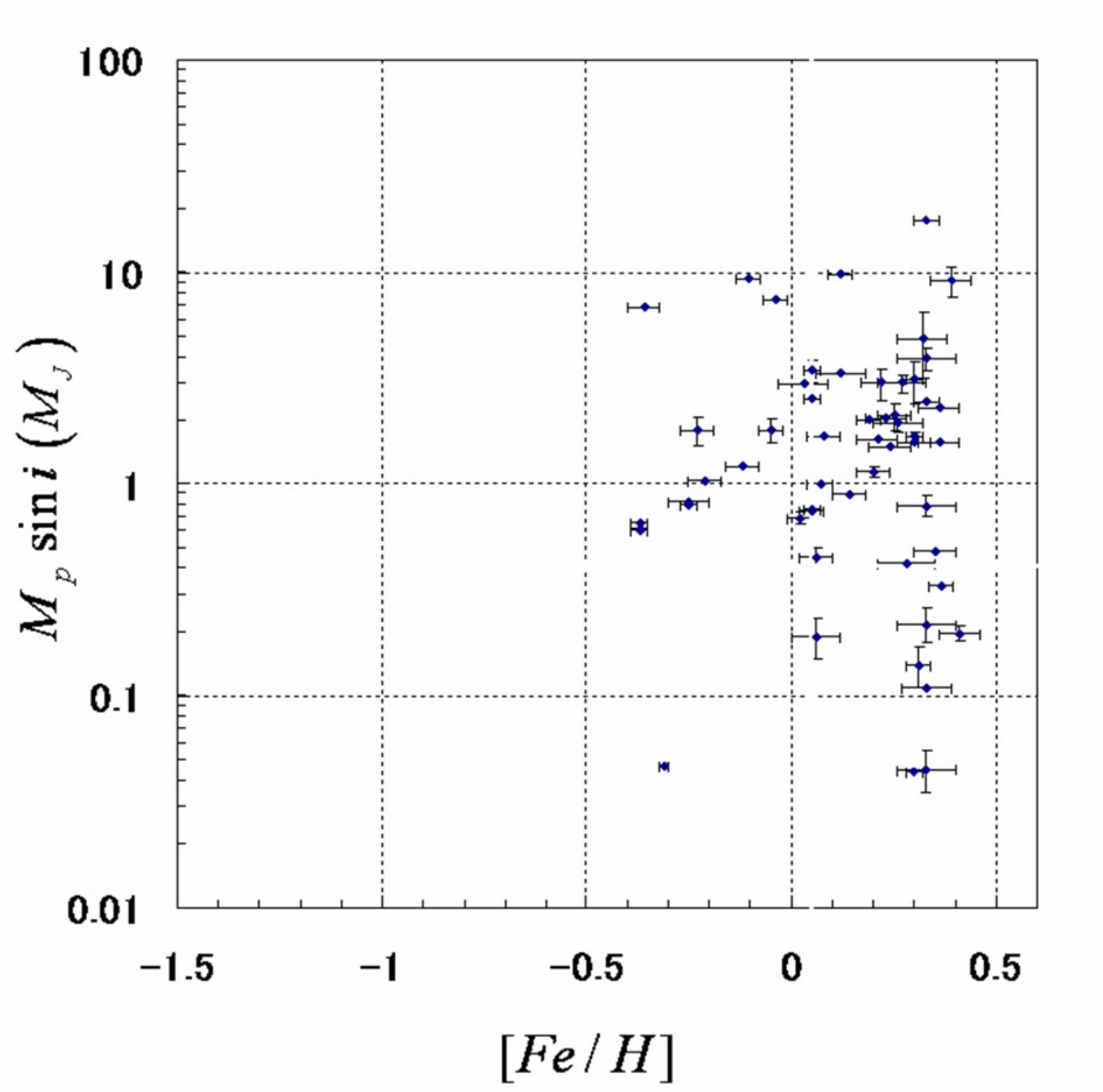

Figure 6c: For G-type dwarfs. (60 samples)

Figure 7: Comparison of samples with theoretical boundaries for the two planetary formation models. The horizontal and vertical axes represent metallicity of the host stars and planet mass, respectively. The solid and dashed lines represent the theoretical boundaries in core-accretion and disk instability models, respectively.

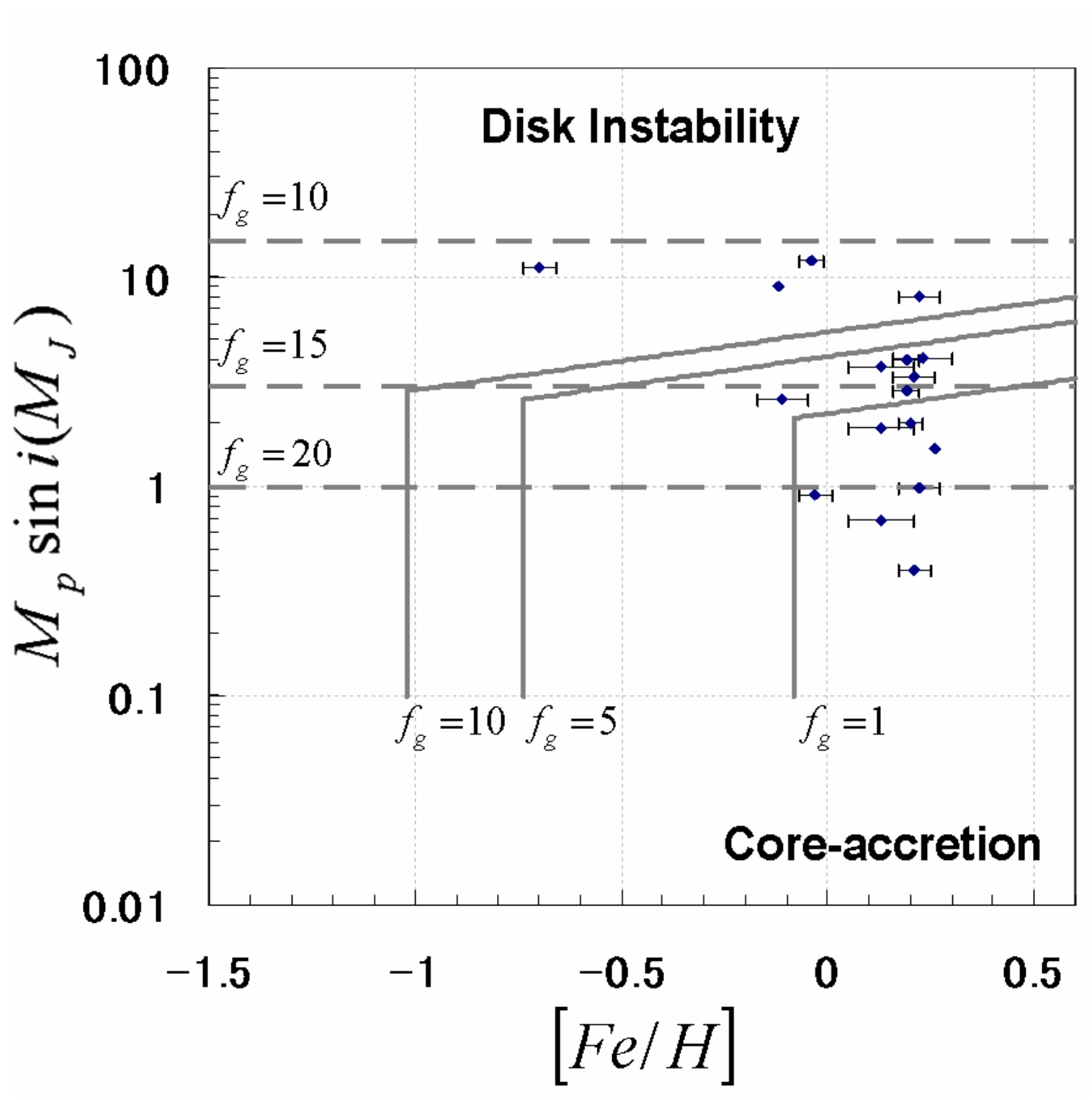

Figure 7a: For F-type dwarfs

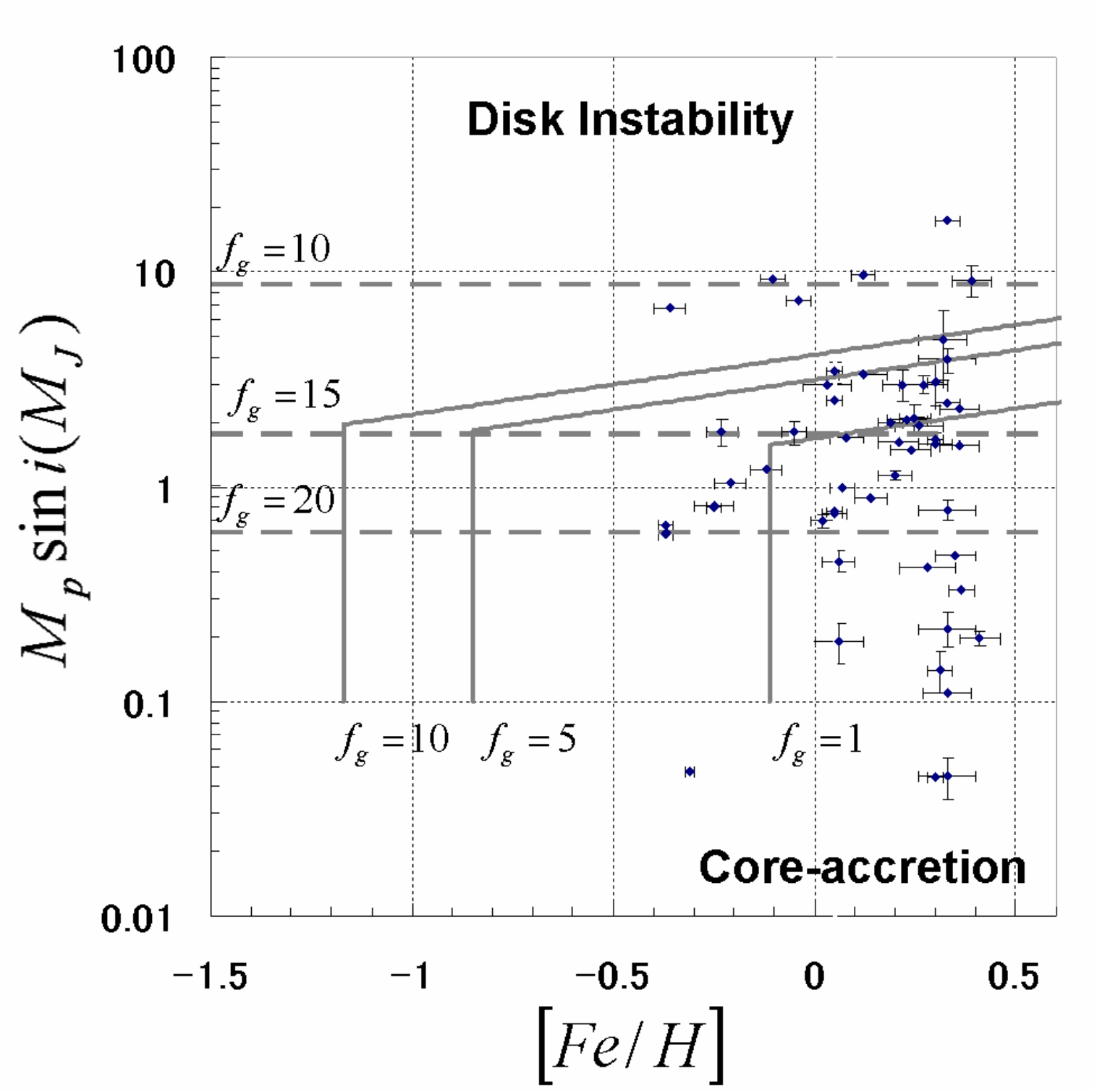

Figure 7b: For G-type dwarfs

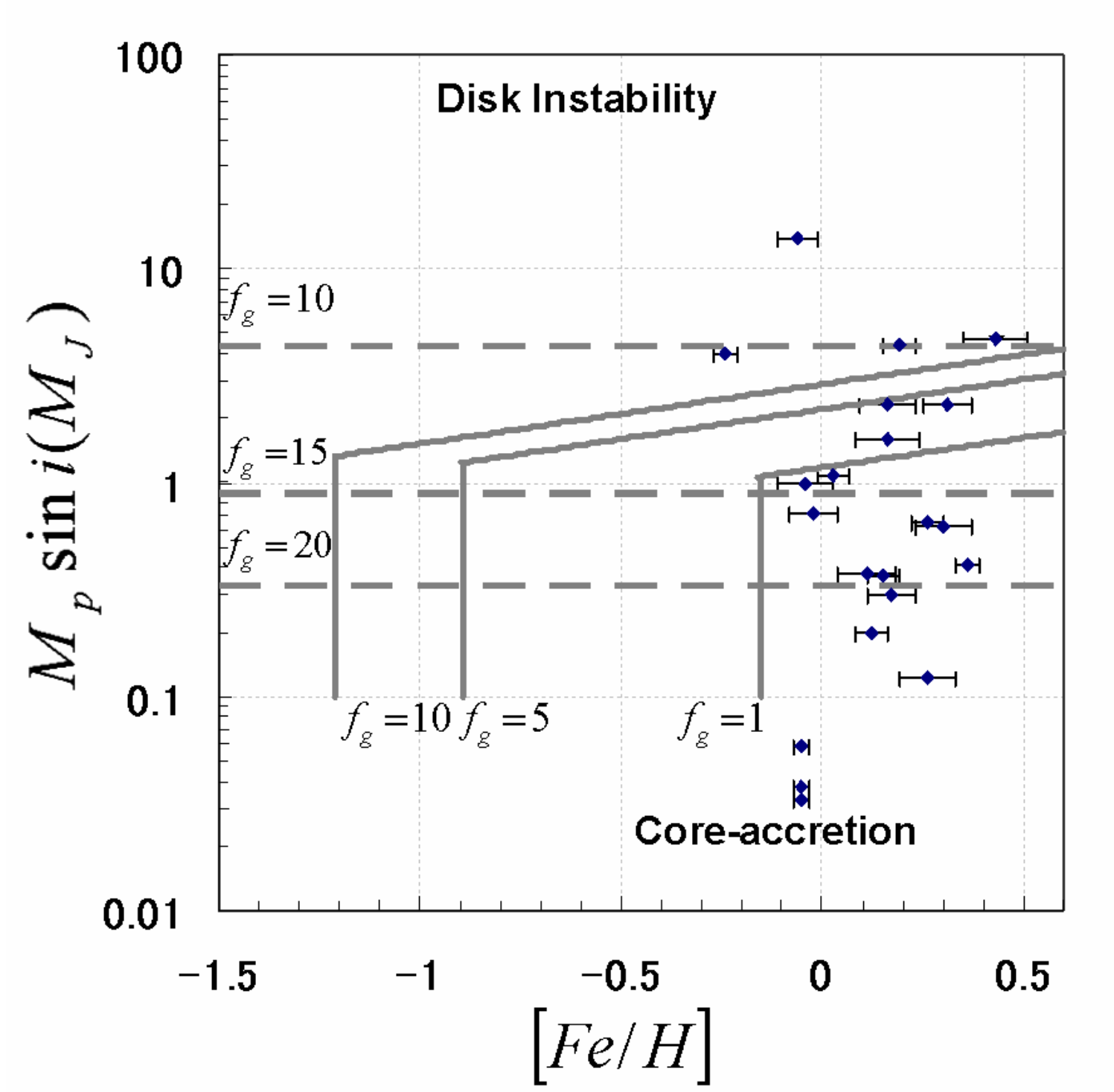

Figure 7c: For K-type dwarfs

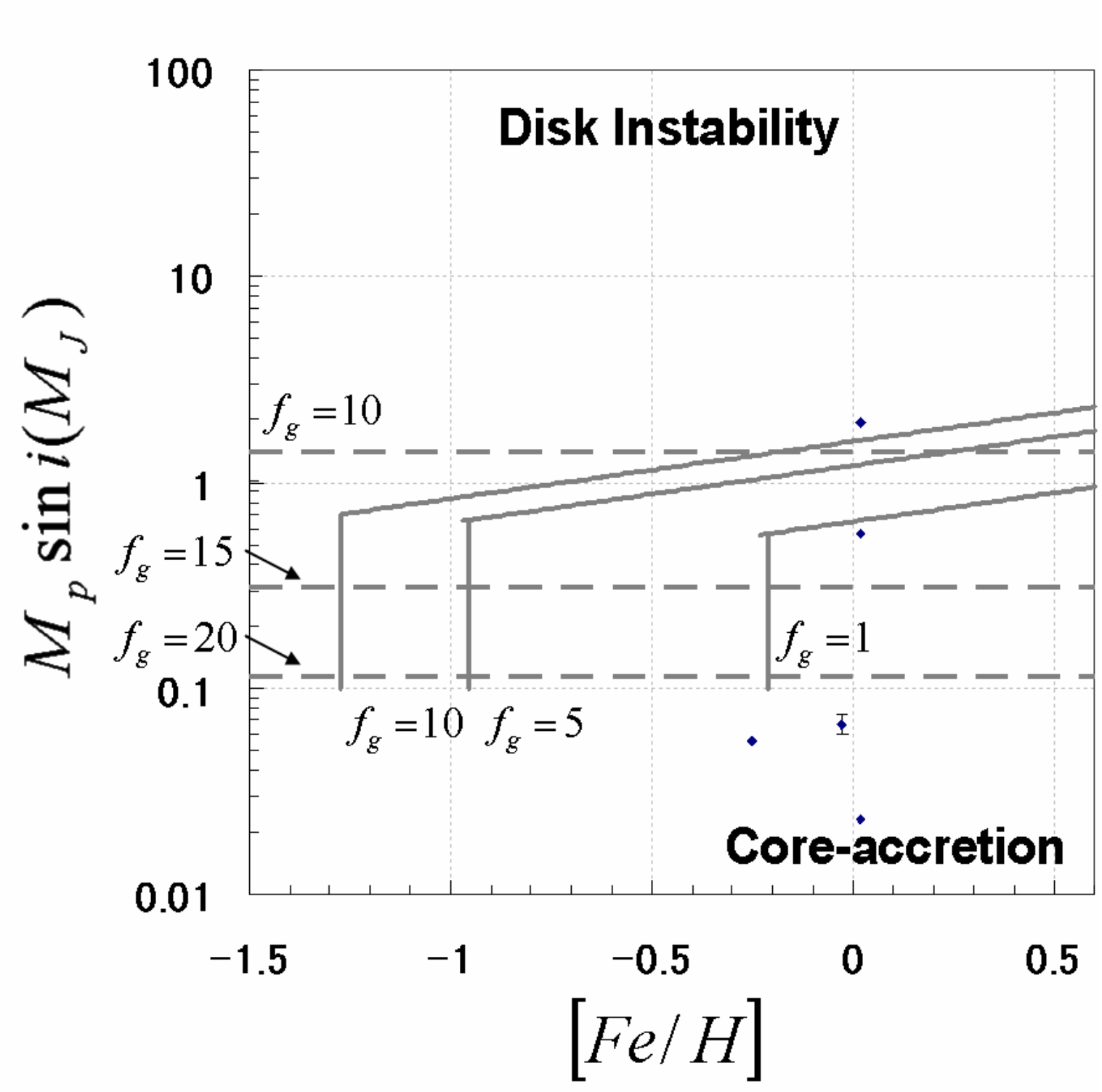

Figure 7d: For M-type dwarfs.

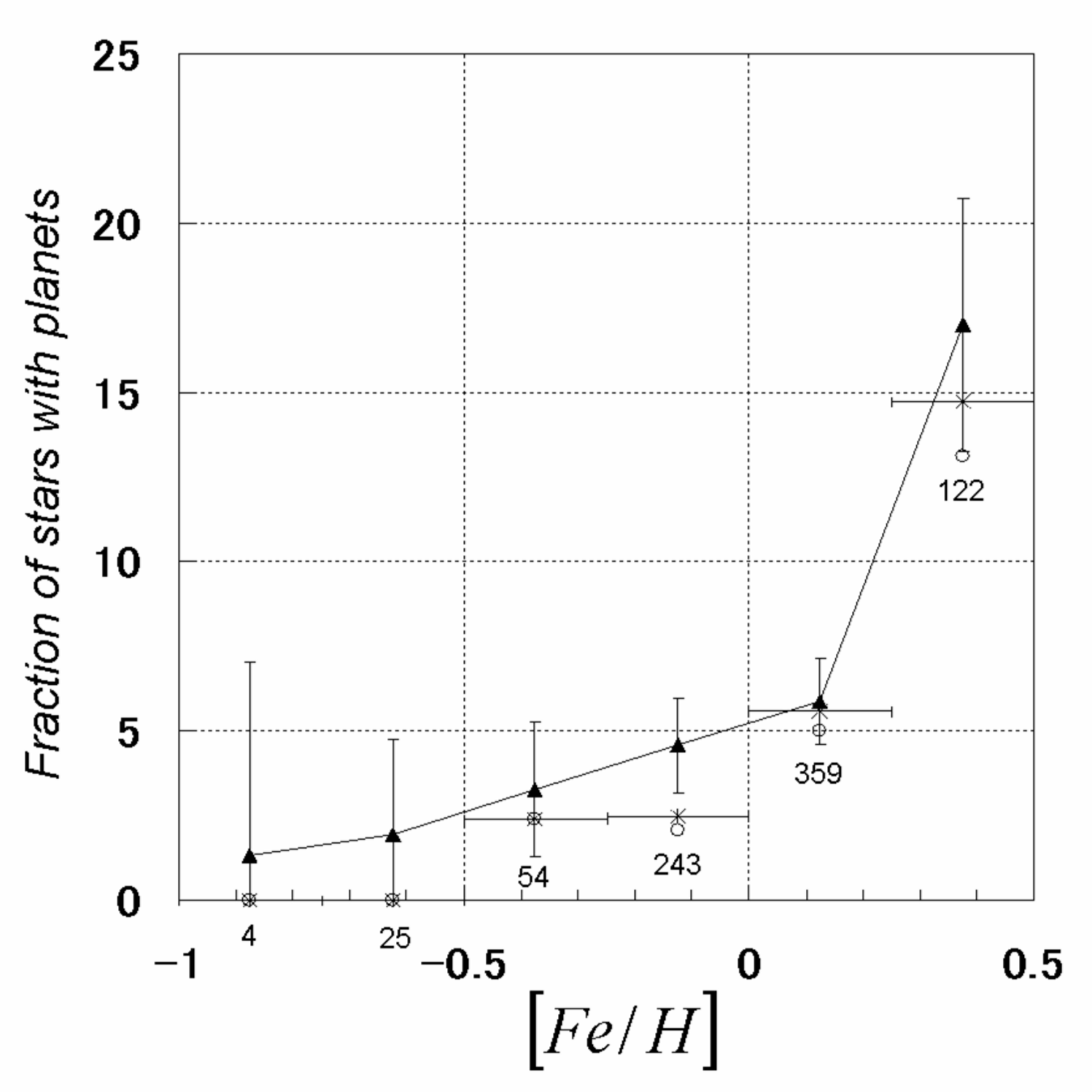

Figure 8: The fraction of stars that harbor gas giants as a function of metallicity. The crosses are the fractions of planet occurrences measured by Fischer & Valenti (2005). Triangles are the percentages of gas giants from the core-accretion model. The reference marks are the fractions of the detected gas giants, and circles are the samples removed from the outer side of the $f_g = 10$ boundary in Figure 7. Each number shows the sample number measured by Fischer & Valenti (2005), and the error bar shows Poisson's error.